\documentclass[a4paper,11pt]{article}
\pdfoutput=1 

\usepackage{jheppub}

\usepackage[ddmmyyyy,hhmmss]{datetime}

\usepackage{amsmath,graphicx,amssymb,subfigure,bbm,comment}
\usepackage[all]{xy}
\usepackage{mathtools}
\usepackage{amsfonts}

\usepackage{xcolor}

\allowdisplaybreaks

\def\be{\begin{equation}}
\def\ee{\end{equation}}
\def\ba{\begin{eqnarray}}
\def\ea{\end{eqnarray}}

\def\a{\alpha}

\def\IR{\relax{\rm I\kern-.18em R}}

\def \ov {\over}

\def\IR{\relax{\rm I\kern-.18em R}}
\def\inv{^{\raise.15ex\hbox{${\scriptscriptstyle -}$}\kern-.05em 1}}

\title{Holographic correlation functions at finite density and/or finite temperature}

\author[a]{George Georgiou}
\author[a,b]{and Dimitrios Zoakos}

\affiliation[a]{Department of Physics, National and Kapodistrian University of Athens, 15784 Athens, Greece}
\affiliation[b]{Department of Engineering and Informatics, 
Hellenic American University, 436 Amherst st, Nashua, NH 03063 USA}

\emailAdd{ggeo@phys.uoa.gr}
\emailAdd{zoakos@gmail.com}

\abstract{We calculate holographically one and two-point functions of scalar operators at finite density and/or finite temperature. In the case of finite density and zero  temperature we argue that only scalar operators can have non-zero VEVs. In the case in which both the chemical potential and the temperature are  finite, we present a systematic expansion of the two-point correlators in powers of the temperature T and the chemical potential $\Omega$.  

The holographic result is in agreement with the general form of the OPE which dictates  that 
the two-point function may be written as a linear combination of the Gegenbauer polynomials $C_J^{(1)}(\xi)$  but with the coefficients depending now on both the temperature and the chemical potential, as well as on the CFT data. The leading terms in this expansion originate from the expectation values of the scalar operator $\phi^2$, the R-current ${\cal J}^\mu_{\phi_3}$ and the energy-momentum tensor 
$T^{\mu\nu}$. 

By employing the Ward identity for the R-current and by comparing the appropriate term of the holographic result for the two-point correlator to the corresponding term in the OPE, we derive the value 
of the R-charge density of the background.  Compelling agreement with the analysis of the thermodynamics of the black hole is found. Finally, we determine the behaviour of the two-point correlators, in the case of finite temperature, and in the limit of large temporal or spatial distance of the operators.}

\begin{document}
\maketitle
\flushbottom


\section{Introduction}

The main scope of the current paper is the study of conformal field theories (CFTs) at finite temperature and finite density 
(chemical potential). In the dual gravity description, the introduction of finite temperature corresponds to the replacement of the 
extremal D3-branes by the non-extremal ones, while the introduction of non-zero R-charges is achieved by generalising the 
static branes to rotating ones.  
This class of solutions was constructed in \cite{Kraus:1998hv}, using previous results from \cite{Cvetic:1996dt}.
They are characterised by a non-extremallity parameter and the rotation parameters, 
which correspond to the three generators of the $SO(6)$ Cartan subalgebra. These rotation parameters
are related to the chemical potentials or equivalently R-charges of the gauge theory. 

More precisely, our main focus will be on the calculation of one and two-point correlation functions between scalar operators 
with large conformal dimension. 
In this limit, the bulk-to-boundary propagator of the dual gravity field can be approximated, in the WKB approximation, by the exponential of minus the mass times the geodesic length of the trajectory for a particle traveling from the boundary point, where the field theory operator is inserted, up to the point in the bulk. Thus, the computation in this limit boils down to the determination of the geodesic.
Even if such a framework looks simplifying, the computation of the two-point function reveals a very interesting structure and 
allow us to probe the physical properties of the dual CFT in the presence of both finite temperature and finite density, or equivalently finite chemical potential. 

A lot of effort has been put so far in the calculation of two-point functions both in the absence (see for example \cite{Tsuji:2006zn,Georgiou:2010an,Georgiou:2011qk}) and in the presence 
of temperature \cite{Balasubramanian:1999zv,Louko:2000tp}.\footnote{For the holographic calculation in backgrounds that are not asymptotically $AdS$ see \cite{Fuertes:2009ex,Georgiou:2018zkt,Park:2022mxj}.} Thermal two-point functions contain much more information than the zero temperature counterparts, 
due to their intricate structure. In the regime where the boundary distance between the operators is much smaller than the size 
of the thermal circle, the operator product expansion (OPE) can be used. Moreover, the presence of the thermal circle provides both a scale and a direction, 
and operators may develop  a vacuum expectation value (VEV). This is the line of reasoning behind the rich structure of the thermal two-point function.
In \cite{Iliesiu:2018fao,Katz:2014rla,Witczak-Krempa:2015pia} general features of the thermal two-point functions have been discussed and it was argued that
they should admit an expansion in terms of Gegenbauer polynomials. The coefficients of this expansion are related to microscopic 
details of the CFT, namely the structure constants of the OPE and the VEVs of operators. Very recently in \cite{Rodriguez-Gomez:2021pfh, Rodriguez-Gomez:2021mkk} (see also \cite{Krishna:2021fus}) thermal two-point functions have been studied in the 
framework of a holographic CFT, using the geodesic approximation. A systematic expansion of the two-point correlator was 
found in terms of the expected Gegenbauer polynomials. The coefficient in the aforementioned expansion, corresponding  
to the energy-momentum tensor was compared successfully with preceding gravity computations. 
Finally the leading terms in the expansion, under a large conformal dimension limit, were resummed into an exponential 
of the block of the energy-momentum tensor. 

Geodesics in black hole backgrounds and their connection with thermal correlations functions has been studied thoroughly 
in the past, see e.g. \cite{Fidkowski:2003nf, Festuccia:2005pi, Hubeny:2006yu, Hubeny:2012ry}. 
More specifically, thermal two-point functions may probe the region behind the horizon of the black hole. This direction has been revived recently 
in \cite{Grinberg:2020fdj} for the one-point function and in 
\cite{Rodriguez-Gomez:2021pfh} for the two-point function case. In \cite{Grinberg:2020fdj} it was argued that thermal 
VEVs may encode the proper time of a radially in-falling particle, from the horizon to the black-hole singularity. 
This is implemented through the presence of higher curvature couplings (e.g. the square of the Weyl tensor $W^2$) which become very large at the vicinity of the singularity. This is the main reason why thermal one-point functions encode the proper time from the horizon to the singularity.
In \cite{Rodriguez-Gomez:2021pfh}  it was argued that a similar mechanism, but now for the thermal two-point function, may also probe the interior of the black-hole.

In this paper, we will extend the holographic study of one and two-point correlation functions using the geodesic approximation, 
to field theories that are both at finite temperature and/or at finite R-charge (chemical potential). 

In section \ref{section:backgrounds}, we will present the gravity backgrounds that we will use for the holographic computation and we will explicitly describe the general method 
for calculating one and two-point functions. This method resembles the one presented in \cite{Rodriguez-Gomez:2021pfh}, enriched with ideas coming from \cite{Janik:2010gc}. 
However,  it is more general, in the sense that it can be applied to geometries in which the components of the metric along the internal coordinates depend on the holographic coordinate. 

In section \ref{section:OPE}, we elaborate on the general structure of the OPE of two scalar operators in the presence of finite density and/or finite temperature, as well as on its implications for the structure of two-point correlation functions. Furthermore, based on the results for the two-point function of section \ref{section:R-charge-T}, as well as on the general form of the OPE presented in section \ref{section:OPE} we calculate the R-charge density of the background and find compelling agreement with the thermodynamics of the black hole.

In section \ref{section:R-charge}, we apply the general method of section \ref{section:backgrounds} to the simple case 
of a gravity background that contains  only R-charge density. Specifically,  we perform the 
holographic calculation of the one and two-point correlation functions of scalar operators of large dimension $\Delta$. 

In section \ref{section:Finite-T}, we review the discussion of \cite{Rodriguez-Gomez:2021pfh} (see also \cite{Rodriguez-Gomez:2021mkk}) for a background that contains only temperature and discuss some issues concerning the large boundary 
distance behaviour of the two-point correlation function. Section \ref{section:R-charge-T} contains the most important results of 
the gravity computation. Combining the effects of the finite temperature and of the finite R-charge, we calculate the one\footnote{In the case of free field theories one-point functions, expressed as single-valued polylogarithms, in the presence of chemical potential and temperature were considered in \cite{Petkou:2021zhg}.} 
and the two-point functions as an expansion for small boundary distance between the operators. 
Equivalently, we present a systematic expansion of the two-point correlators in powers of the temperature T and the R-charge chemical potential $\Omega$. 


\section{Backgrounds and set-up for calculating correlation functions}
\label{section:backgrounds}

Our starting point will be the the non-extremal rotating D3-brane solution of  
type-IIB supergravity that was first found in \cite{Russo:1998mm,Kraus:1998hv} (based on \cite{Cvetic:1996dt}). 
Its metric reads
\begin{eqnarray}  \label{metric-general-v1}
ds^2 &=& H^{-\frac{1}{2}}\Bigg[-\left(1-\eta^4 \, H \right) dt^2 + d\vec{x}_3^2 \Bigg]  + H^{\frac{1}{2}}
\Bigg[{dz^2\over f}  + \frac{\tilde \Delta}{z^2} \, d\theta^2 + \frac{1}{z^2} \, \cos^2\theta \, d\Omega_3^2 
\nonumber\\
&& + \, \frac{1}{z^2} \, \left(1-r_0^2 \, z^2\right) \sin^2\theta\ d\phi_1^2  
- 2 \, \eta^2 \, r_0 \,   \cos^2\theta  \, dt \left(\sin^2 \psi \, d\phi_2+ \cos^2 \psi \, d\phi_3\right) \Bigg] 
\end{eqnarray}
where
\begin{equation} \label{def-functions-general}
H  =  {z^4  \over \tilde \Delta}\ , \quad 
f =  H \, \left(1 - \eta^4 \, z^4 \right) \, \left(1 - r_0^2 \, z^2 \right)  \quad {\rm with} \quad 
\tilde \Delta = 1- r_0^2 \, z^2 \cos^2\theta\, .
\end{equation}
Furthermore, the line element of the three-sphere is given by
\begin{equation}
d\Omega_3^2 \, = \, d\psi^2 + \sin^2 \psi \, d\phi_2^2+ \cos^2 \psi \, d\phi_3^2 \, . 
\end{equation}
In the above expressions, $r_0$ denotes the angular momentum parameter 
and $\eta$ is related to the temperature. The location of the horizon is at the $z_H = \eta^{-1}$. The various 
thermodynamic quantities were calculated in \cite{Gubser:1998jb,Harmark:1999xt,Russo:1998by} 
(see also \cite{Avramis:2006ip}) and read
\begin{equation} \label{thermo-quantities}
T= \frac{1}{\pi} \, \sqrt{\eta^2 - r_0^2}\, , \quad 
S= \frac{1}{2\, \pi} \, N^2 \, \eta^2  \sqrt{\eta^2 - r_0^2} \, , \quad 
\Omega = r_0 \quad \& \quad J = \frac{1}{4\, \pi^2} \, r_0 \, \eta^2 \, N^2 \, . 
\end{equation}
The only constraint that needs to be imposed is the reality condition \( \eta \ge r_0 \).
It should be emphasized that in order to extract information about the gauge theory, the supergravity parameters 
(\(\eta \, \& \, r_0\)) should be traded either with (\( T \, \& \, J \)) or with (\( T \, \& \, \Omega \)), 
where $J$ and $\Omega$
play the role of R-charge density and chemical potential, respectively. Notice that the choice of (\(T \, \& \, J\)) 
corresponds to the Canonical Ensemble  while the choice of (\( T \, \& \, \Omega\)) to the Grand Canonical Ensemble.

Our analysis will focus on two point-like string solutions which sit either at ($\theta=0$ \& $\psi=0$) or at 
($\theta=\pi/2$ \& $\psi=0$). 
One can easily check that this is consistent with the full 10-dimensional set of equations of motion. 
For $\theta=\pi/2$ \& $\psi=0$, the metric \eqref{metric-general-v1} effectively becomes
\begin{equation} \label{metric-general-theta-pi-over-2}
ds^2 = \frac{1}{z^2} \Bigg[- f_1 \, dt^2 + d\vec{x}_3^2 +\frac{ dz^2}{g\, f_1} \Bigg] + g \, d\phi_1^2
\,\,\,\, {\rm with} \,\,\,\,
f_1 = 1 - \eta^4 \, z^4
\,\,\,\, \&  \,\,\,\,g = 1-r_0^2 \, z^2 \, ,
\end{equation}
while for $\theta=0$ \& $\psi=0$, the metric \eqref{metric-general-v1} takes the form
\begin{equation} \label{metric-general-theta-zero}
ds^2 = \frac{1}{z^2} \Bigg[-f_2 \, dt^2 +  \sqrt{g} \, d\vec{x}_3^2 +\frac{dz^2 }{\sqrt{g}\, f_1} \Bigg] + \frac{ d\phi_3^2}{\sqrt{g}} - 
2 \, \frac{r_0 \, \eta^2 \, z^2}{\sqrt{g}} \, dt \, d\phi_3
\,\,\, {\rm with} \,\,\,
f_2 = \sqrt{g} - \frac{\eta^4 \, z^4}{\sqrt{g}} \, . 
\end{equation}
At this  point, let us mention that background \eqref{metric-general-v1} is dual to ${\cal N}=4$ SYM at finite density (chemical potential) and at finite temperature.

In what follows, we will present our general method for calculating one and two-point correlation functions of operators 
that are dual to classical string states at strong coupling. The analysis is for point-like strings, that is for string 
solutions without $\sigma$ dependence.  As mentioned above, it is consistent with the 10-dimensional equations of motion to keep only one of the sphere isometries. In such a case the metric assumes the following generic form  
\begin{equation} \label{metric-general-g2}
ds^2 = g_{tt}(z) \, dt^2 + g_{xx}(z)\, d\vec{x}_3^2 +g_{zz}(z) \, dz^2 +g_{\phi\phi}(z) \, d\phi^2+
2 g_{t\phi}(z)dt\, d\phi \,  .
\end{equation}
The holographic calculation proceeds as follows. Firstly, one writes down the Polyakov action for the string $S_P$ and then performs a Legendre transformation with respect to the isometry $\phi$, in order to pass from the variables 
$(\phi,\dot\phi)$ to  $(\phi,\cal\pi_\phi)$ where $\cal\pi_\phi$ is the momentum conjugate to the isometry $\phi$, which is a constant of the motion.\footnote{In the general case, one should perform such Legendre transforms along all the isometries of the background.} 
This Legendre transform implements  the convolution of the semiclassical
propagator with the wavefunction of the state that we are interested in \cite{Janik:2010gc}.  
Subsequently, one solves the non-trivial Virasoro constraint $g_{\mu\nu} \partial_\tau X^\mu \partial_\tau X^\nu=0$ for $\dot\phi$ or equivalently for $\pi_\phi\equiv \Delta$ and substitutes the result in the action. In this way one ends up with the following action\footnote{Notice that we have kept only one of the spatial directions of $\vec{x}_3$ in 
\eqref{metric-general-g2}.}
\begin{equation}  \label{NG}
 S=S_P-\int d^2 \sigma \,\pi_\phi \dot\phi=- \Delta \int \frac{d \tau}{g_{\phi\phi}}
\Bigg[ -g_{t\phi} \, \dot t+\sqrt{\Big.g_{t\phi}^2 \, \dot t^2-g_{\phi\phi} 
\left(g_{zz} \, \dot z^2+g_{tt} \, \dot t^2+g_{xx} \, \dot x^2 \right)}\Bigg] \, .
\end{equation}

The virtue of \eqref{NG} is that we have integrated out the coordinates of the internal space. In this way 
the effective action \eqref{NG} describes the motion of a point-like string in the 5-dimensional space 
$(t,z,\vec x)$ with the motion of the particle in the sphere taken into account correctly.
One may now choose the gauge $z=\tau$ and solve the equations of motion for the isometries $t$ and $x$ to get
\begin{equation}   \label{eom1}
\pi_t \, = \, \frac{\partial \cal L}{\partial \dot t} \, = \, \mu \,\Delta  
\quad \& \quad 
\pi_x \, = \, \frac{\partial \cal L}{\partial \dot x} \, = \, \nu \,\Delta
\end{equation}
where $\mu$ and $\nu$ are constants. The above equations can now be solved for $\dot t$ and $\dot x$ to give
\begin{equation}\label{tdxd}
\dot t \, = \, i \, \sqrt{\big. g_{zz} \, g_{xx}}\frac{g_{t\phi} -\mu \,g_{\phi\phi}}{D} \quad  \& \quad 
\dot x \, = \, i \, \nu \sqrt{\frac{g_{zz}}{g_{xx}}} \, 
\frac{g_{t\phi}^2 - g_{\phi \phi} g_{tt} }{D}
\end{equation}
where $D$ is defined as follows
\begin{equation}
D^2 \, = \, \left(g_{t\phi}^2 - g_{\phi \phi} g_{tt} \right)\, \Big[-g_{xx}\big[g_{tt}+\mu 
\left(\mu\, g_{\phi \phi}-2 \, g_{t\phi}\right) \big]+ \nu^2 \left(g_{t\phi}^2 - g_{\phi \phi} \, g_{tt}\right)\Big]\, .
\end{equation}
Substituting \eqref{tdxd} in  \eqref{NG} one gets the final expression for the action functional
\begin{equation} \label{NGfin}
S \, = \, i \, \Delta \int dz \sqrt{\big.g_{zz} \, g_{xx}}  \, \frac{g_{tt}-\mu \, g_{t \phi}}{D} \, .
\end{equation}
This is the expression that we will be using for the holographic calculation of the correlation functions in the 
backgrounds of interest. 

In order to work with Euclidean signature we perform the following analytic continuation  to the metrics 
\eqref{metric-general-theta-pi-over-2} and \eqref{metric-general-theta-zero}
\begin{equation}
t \rightarrow i \, t \quad \& \quad \phi \rightarrow - \, i \, \phi \, . 
\end{equation}
As a result the metric components that will enter the general formulas for the $\theta=\pi/2$ \& $\psi=0$ case will be 
\begin{equation} \label{metric_comp_theta_pi_over_2}
g_{tt} = \frac{ f_1}{z^2} \, , \quad
g_{xx} = \frac{1}{z^2} \, , \quad
g_{zz} = \frac{1}{z^2}\frac{1}{g \, f_1}  \quad \& \quad 
g_{\phi \phi} = - \, g 
\end{equation}
while for the $\theta=0$ \& $\psi=0$ case will be 
\begin{equation} \label{metric_comp_theta_zero}
g_{tt} = \frac{f_2}{z^2} \, , \quad
g_{xx} = \frac{\sqrt{g}}{z^2} \, , \quad
g_{zz} = \frac{1}{z^2}\frac{1}{\sqrt{g} \, f_1}  \, , \quad
g_{\phi \phi} = - \, \frac{1}{\sqrt{g}} \quad \& \quad 
g_{t \phi} = - \, \frac{r_0 \, \eta^2 \, z^2}{\sqrt{g}} \, . 
\end{equation}
For the case of zero density $r_0=0$ our expression \eqref{NGfin} reduces to
\begin{equation} \label{auxiliary}
S \, = \, \Delta  \int \frac{dz}{z} \frac{1}{\sqrt{\big.\left(1-\eta ^4 \, z^4\right) \left(1-\nu^2 \, z^2\right) -\mu^2 \, z^2}}
\end{equation}
that agrees perfectly with the one obtained in \cite{Rodriguez-Gomez:2021pfh}. 

The virtue of our approach, in which the effective action is obtained by using the Polyakov instead of the Nambu-Goto action, is that it can be applied to the holographic calculation of correlation functions in more general backgrounds in which the components of the metric along the internal coordinates depend on the holographic coordinate. In a sense, the effective action \eqref{NG} takes correctly into account the motion of the particle in the internal space. Notice that the internal directions do not appear in the action  \eqref{NG}. Notice, also, that action  \eqref{NG} gives the same result with the Nambu-Goto action used in \cite{Rodriguez-Gomez:2021pfh,Rodriguez-Gomez:2021mkk} only when the 10-dimensional metric is a product of a 5-dimensional space and an undeformed 5-sphere of constant radius, as it happens in the case of zero density and finite temperature.


\section{OPE and correlation functions}
\label{section:OPE}

In this section, we elaborate on the operator product expansion of two scalar operators and its implications for the structure of two-point correlation functions. We start with the case of finite density and zero  temperature. Subsequently, we consider the case where both density and temperature are non-zero. In the latter case, we use the two-point correlation function calculated holographically  in subsection \ref{subsubsection:Finite-T-R-temporal-spatial} together with the general structure of the OPE and the Ward identity which the R-current obeys in order to derive the R-charge density. Our expression is in complete agreement with the one obtained from the type-IIB supergravity solution.


\subsection{Finite density and zero temperature}

The OPE of two scalar operators takes the generic form
\be\label{OPE}
 {\cal O}(x) {\cal O}(0) =\sum_{\tilde {\cal O}\in {\cal O} \times {\cal O}} \frac{f_{\cal O\cal O \tilde {\cal O}}}{c_{ \tilde {\cal O}}}
 \frac{|x|^{\Delta_{\tilde {\cal O}}}}{|x|^{2\Delta_{\cal O}+J}} x_{\mu_1} \cdots x_{\mu_J} \tilde {\cal O}^{\mu_1 \cdots \mu_J}(0)
 \ee
where $f_{\cal O\cal O \tilde {\cal O}}$ is the three-point coefficient that controls the following three-point correlator, namely
\be\label{OPE-3-point}
\langle  {\cal O}(x_1) {\cal O}(x_2) \tilde {\cal O}(x_3)\rangle=f_{\cal O\cal O \tilde {\cal O}}\frac{Z^{\mu_1}\cdots Z^{\mu_J}-{\rm traces}}{x_{12}^{2\Delta_{\cal O}-\Delta_{\tilde {\cal O}}+J} 
x_{13}^{\Delta_{\tilde {\cal O}}-J}x_{23}^{\Delta_{\tilde {\cal O}}-J}}
\quad \& \quad Z^{\mu}=\frac{x_{13}^\mu}{x_{13}^2}-\frac{x_{23}^\mu}{x_{23}^2} \, . 
\ee
Furthermore, $c_{\cal O}$ is the normalisation constant of the two-point correlator
\begin{equation}\label{2-point}
\langle  {\cal O}^{\mu_1\cdots \mu_J}(x) {\cal O}_{\nu_1\cdots \nu_J}(0) \rangle=c_{\cal O} \frac{I^{(\mu_1}_{(\nu_1} \cdots I^{\mu_J)}_{\nu_J)}-{\rm traces}}{x^{2\Delta_{\tilde {\cal O}}}}
\quad {\rm with} \quad I^{\mu}_{\nu}=\delta^{\mu}_{\nu}-2\frac{x^{\mu}x_{\nu}}{x^2} \, .
\end{equation}
As usual, one may normalise $\cal O$ so that $c_{\cal O}=1$.

One may now take the expectation value of  \eqref{OPE} to obtain the two-point correlation function of scalar operators as an infinite sum of one-point functions.\footnote{In the conformal case in which the dual gravity background is $AdS$ only one operator, namely the identity, contributes in the right hand side of \eqref{OPE}.}
An important comment is in order.
Notice that because the descendants have vanishing one-point functions, one needs only the leading (non-derivative) term in the OPE for each multiplet.

In the case of zero temperature the dual field theory lives on $\mathbb{R}^4$, as can be easily seen from \eqref{metric-general-v1}. Translational invariance combined with dimensional analysis, that is with the fact that for $\eta=0$,  the geometries
 \eqref{metric-general-theta-pi-over-2} and  \eqref{metric-general-theta-zero} have a scale, namely $r_0$, implies that all operators with spin have zero one-point functions. Only scalar operators may have non-zero one-point functions in the background of zero temperature and finite density, namely
\be\label{1-point}
\langle  {\cal O}^{\mu_1\cdots \mu_J}(x)\rangle_{\mathbb{R}^4}=0, \qquad \langle  {\cal O}(x)\rangle_{\mathbb{R}^4}= a_{\cal O}\, r_0^{\Delta_{\cal O}}\, .
\ee
We will see, that this result is in perfect agreement with the holographic calculation of the two-point functions of 
scalar operators in the presence of finite density (chemical potential) at zero temperature 
(see section \ref{subsection-R-charge-2point}).
In particular, it explains why there is no $|x|^3$ term in the exponent of the two-point correlators 
\eqref{R-2point-theta-pi-over2-v2} and \eqref{R-2point-theta-zero-v2} . It is so because the R-currents $j^{\mu}_R$ which are associated to the isometries of the background and which have dimension $\Delta_{\tilde {\cal O}}=\Delta_{j^{\mu}_R}=3$ have zero one-point function and as a result they do not contribute to the right hand side of \eqref{OPE}.
Finally, plugging \eqref{1-point} in \eqref{OPE} one obtains
\be\label{2-point-R}
\langle {\cal O}(x) {\cal O}(0) \rangle=\sum_{\tilde {\cal O}\in {\cal O} \times {\cal O}} \frac{f_{\cal O\cal O \tilde {\cal O}}}{c_{ \tilde {\cal O}}}\frac{ a_{\tilde {\cal O}}}{|x|^{2\Delta_{\cal O}-\Delta_{\tilde {\cal O}}}} 
 \, r_0^{\Delta_{\tilde {\cal O}}}\, .
 \ee
This generic expression for the two-point function is in complete agreement with the result obtained from the holographic calculation of the two-point correlators, see \eqref{R-2point-theta-pi-over2-v2} and \eqref{R-2point-theta-zero-v2}. 

\subsection{Finite density and non-zero temperature}
In the case of both non-zero density and temperature \eqref{OPE} implies
\be\label{OPE-1}
\langle {\cal O}(x) {\cal O}(0)\rangle_{\beta, r_0} =\sum_{\tilde {\cal O}\in {\cal O} \times {\cal O}} \frac{f_{\cal O\cal O \tilde {\cal O}}}{c_{ \tilde {\cal O}}}\frac{|x|^{\Delta_{\tilde {\cal O}}}}{|x|^{2\Delta_{\cal O}+J}} x_{\mu_1} \cdots x_{\mu_J}
 \langle \tilde {\cal O}^{\mu_1 \cdots \mu_J}(0)\rangle_{\beta, r_0}.
 \ee
In this case the time direction can be distinguished and operators with spin can have non-zero expectation values, namely
\be\label{1-point-2}
\langle  {\cal O}^{\mu_1\cdots \mu_J}(x)\rangle_{\beta,r_0}=\frac{1}{\beta^{\Delta_{\cal O}}}
\sum_{n=0}^{\Delta_{\cal O}}a^{(n)}_{\cal O}(r_0 \, \beta)^n\big(e^{\mu_1}\cdots e^{\mu_J} -{\rm traces} \big) 
\quad {\rm where}\quad  e^{\mu}=\delta ^{\mu\,t} \, . 
\ee
Two important operators with spin which can appear in the right-hand side of \eqref{1-point-2} are the following 
conserved quantities, namely the R-currents $j^{\mu}_R$ and the stress-energy tensor $T^{\mu \nu}$ of the dual field theory. Plugging now \eqref{1-point-2} in \eqref{OPE-1} and performing the tensor contractions one obtains the following result
\be\label{2-point-RT}
\langle {\cal O}(x) {\cal O}(0) \rangle=\sum_{\tilde {\cal O}\in {\cal O} \times {\cal O}}\sum_{n=0}^{\Delta_{\tilde {\cal O}}} \frac{f_{\cal O\cal O \tilde {\cal O}}}{c_{ \tilde {\cal O}}}\frac{ a^{(n)}_{\tilde {\cal O}}(r_0 \, \beta)^n}{|x|^{2\Delta_{\cal O}-\Delta_{\tilde {\cal O}}}} 
 \, \frac{1}{\beta^{\Delta_{\tilde {\cal O}}}}\frac{1}{2^{J_{\tilde {\cal O}}}} C_{J_{\tilde {\cal O}}}^{(1)}(\xi)
\quad {\rm where} \quad \xi=\frac{\tau}{|x|} 
 \ee
and $C_{J_{\tilde {\cal O}}}^{(1)}(\xi)$ are the Gegenbauer polynomials. Each of the terms in the sum is originating from an operator of dimension $\Delta_{\tilde {\cal O}}$ and spin $J_{\tilde {\cal O}}$.
 
In the presence of both chemical potential and temperature we will distinguish two cases. The first is when the string solution dual to the field theory operators sits at $\theta=\frac{\pi}{2}$. In this case, and similar to what happens in the case of zero temperature, there is no $|x|^3$ term (the term associated to the R-current) in the exponent of the two-point correlator, see equations  \eqref{RT-full-2point-theta-pi-over2_v1} and \eqref{RT-full-2point-theta-pi-over2_v3}. 
 The explanation of this fact is the following: In contradistinction to the case of zero temperature, two components of the R-current operator, the one associated with $\phi_3$ and the one associated with $\phi_2$, acquire non-zero expectation value. In our calculations we have chosen $\psi=0$ which implies that only the R-current associated to $\phi_3$ survives. The corresponding operator is the conserved current which can be derived from the fact that 
 $\phi_3$ is an isometry of the background, namely $j^{\mu}_{\phi_3}$. However, one can easily see that the operator sitting at $\theta=\frac{\pi}{2}$ is not charged under the $\phi_3$ isometry. This in turn implies that the three-point coefficient $f_{{\cal O} {\cal O} j^{\mu}_{\phi_3}}$ appearing in the right hand side of \eqref{OPE-1} is zero, since there is a Ward identity which states that the aforementioned OPE coefficient is proportional to the charge of the operator under the symmetry generating the current. This explains  the absence of the $|x|^3$ term in the exponent of the two-point correlator in 
 \eqref{RT-full-2point-theta-pi-over2_v1}, despite the fact that $\langle j^{\mu}_{\phi_3} \rangle\neq 0$.
 
The second case is when the string solution dual to the field theory operators sits at $\theta=0$. In this case, there exists a term proportional to $|x|^3$ term in the exponent of the two-point correlator in \eqref{RT-full-2point-theta-zero_v1} (see also \eqref{RT-full-2point-theta-zero_v3}),
 since in this case the scalar operator is charged under the isometry $\phi_3$. As a result the three-point function coefficient is not zero, i.e. $f_{{\cal O} {\cal O} j^{\mu}_{\phi_3}} \neq 0$. In fact, employing the calculation the two-point correlator presented in subsection \ref{subsubsection:Finite-T-R-temporal-spatial}, one can extract a prediction for the one-point of the R-current $j^{\mu}_{\phi_3}$ at the strong coupling regime. This will be done in next subsection.


 \subsection{R-charge density and agreement with the SUGRA solution}

We start by keeping in \eqref{OPE-1} only the term that is relevant for our purposes, namely
\be\label{OPE-2}
\langle {\cal O}(x) {\cal O}(0)\rangle_{\beta, r_0} =\cdots+  \frac{f_{{\cal O}{\cal O} J_{\phi_3}^{\mu}}}{c_{ {\cal J}_{\phi_3}^{\mu}}}\frac{|x|^{3}}{|x|^{2\Delta_{\cal O}}} \frac{x_{\mu} }{|x|}
 \langle  {\cal J}_{\phi_3}^{\mu}(0)\rangle_{\beta, r_0}.
 \ee
 By comparing \eqref{OPE-2} to \eqref{RT-full-2point-theta-zero_v1} we deduce that
 \be\label{OPE-3}
 \frac{f_{{\cal O}{\cal O} J_{\phi_3}^{\mu}}}{c_{ {\cal J}_{\phi_3}^{\mu}}} \langle  {\cal J}_{\phi_3}^{0}(0)\rangle_{\beta, r_0}
 =-{1 \ov 6} \, r_0 \, \eta^2 \, \Delta_{\cal O}\, .
 \ee
In order to find the  density of the R-charge $\langle  {\cal J}_{\phi_3}^{0}(0)\rangle_{\beta, r_0}$ from the holographic 
computation of the two-point function of scalar operators we need the OPE coefficient 
$f_{{\cal O}{\cal O} J_{\phi_3}^{\mu}}$, as well as the normalisation constant for the R-current $c_{ {\cal J}_{\phi_3}^{\mu}}$. The first piece of information can be derived from the Ward identity which the  R-current obeys. The OPE coefficient takes the value \cite{Georgiou:2013ff}
 \be\label{OPE-coeff}
 f_{{\cal O}{\cal O} J_{\phi_3}^{\mu}}=-{J \ov 2 \, \pi^2}\, .
 \ee
In order to find the  normalisation constant for the R-current, one should write down its precise form which reads
\be\label{R-current}
{\cal J}_{A=3}^{ \mu\,\,\,B=4}=\phi_{3C}\overleftrightarrow{\partial}^\mu \phi^{C4}+ \bar \lambda_{\dot \a 3}\bar\sigma^{\mu \dot \a \a}\lambda_{ \a}^{ 4}=\phi_{31}\overleftrightarrow{\partial}^\mu \phi^{14} 
+\phi_{32}\overleftrightarrow{\partial}^\mu \phi^{24}+ \bar \lambda_{\dot \a 3}\bar\sigma^{\mu \dot \a \a}\lambda_{ \a}^{ 4}\, .
\ee
Notice that in the last expression we have suppressed the color indices of the current.
Let us also mention that the scalars in \eqref{R-current} have the canonical two-point function, 
namely  $\langle\phi_{AB}(x)\phi^{AB}(0)\rangle=-{1\ov 4 \, \pi^2 x^2} $.
The normalisation of the R-current can now be found by evaluating the two-point function $\langle\Big({\cal J}_{3}^{ \mu\,4}(x)\Big)^C \,\Big({\cal J}_{4}^{ \mu\,3}(0)\Big)^C\rangle$, where we have restored the colour structure and $C=1,2\cdots,N^2-1$ for the gauge group $SU(N)$. The complete normalisation is as follows\footnote{We work in the large $N$ limit approximating thus $N^2-1\approx N^2$, where $N$ is the number of colours of the dual field theory.}
\be\label{norm}
c_{ {\cal J}_{\phi_3}^{\mu}}=\tilde N \left[ 4 \times {1 \ov 2} \, c_\phi+ c_\lambda\right] \, = \, 3\,  {N^2 \ov 4 \, \pi^4} 
\ee
where we have taken into account that $\tilde N={\rm tr}\left(t_C \, t_D\right)={1 \ov 2} \, \delta_{CD}$ 
while the normalisations $c_\phi={1 \ov 2}{N^2 \ov 4 \pi^4}$ and $ c_\lambda=4 \, {N^2 \ov 4 \pi^4}$ 
can be straightforwardly calculated and can be found in section 5 of \cite{Osborn:1993cr}. 
Finally, the factor of 4 appearing just after the bracket in \eqref{norm} originates from the fact that there are 4 different scalar contributions in the correlator $\langle\Big({\cal J}_{3}^{ \mu\,4}(x)\Big)^C \,\Big({\cal J}_{4}^{ \mu\,3}(0)\Big)^C\rangle$.
This is so because  the scalar part of each of the current consists of two terms.

Plugging \eqref{norm} and \eqref{OPE-coeff} in \eqref{OPE-3} and taking into account that $ \Delta_{\cal O}=J$, one obtains for the R-charge density the following expression
\be\label{density}
\langle  {\cal J}_{\phi_3}^{0}(0)\rangle_{\beta, r_0} \, = \, r_0 \, \eta^2{N^2 \ov 4 \, \pi^4} \, . 
\ee
This result is in complete agreement with the thermodynamics of the black hole solution \eqref{thermo-quantities}, see also for example equation (2.13) of \cite{Avramis:2006ip} (see also \cite{Gubser:1998jb,Russo:1998by}). 
We conclude that the general form of the OPE combined with the holographic result for the two-point correlator of scalar operators gives a prediction for the density of the R-charge, which is in complete agreement with the value obtained from the supergravity solution.



\section{R-charge density}
\label{section:R-charge}

In this section, 
we will present the holographic calculation of the one and two-point correlators in the presence of only the 
R-charge.  We will be able to obtain closed expressions for some of the observables. 
The methodological approach that we follow is similar to the one used in \cite{Grinberg:2020fdj}  for the one-point function,
and similar to the one used in  \cite{Rodriguez-Gomez:2021pfh,Rodriguez-Gomez:2021mkk} for the two-point function. 
Of  course the use of the geodesic arcs in order to compute correlation functions of heavy operators  at strong coupling, 
is an approach that has appeared a long time ago in the literature   
\cite{Fidkowski:2003nf, Festuccia:2005pi, Hubeny:2006yu, Hubeny:2012ry}. .

\subsection{One-point correlation function}
\label{subsection-R-charge-1point}

Initially we focus attention on the calculation of the one-point function of scalar operators. There are two cases to cover. 
In the first one, we will calculate the one-point function for operators dual to a point-like string that sits 
at $\theta=\pi/2$ and $\psi=0$ with the initial metric reducing to \eqref{metric-general-theta-pi-over-2}. Then we will consider the case in which the string  sits at $\theta=0$ and $\psi=0$ with the corresponding metric becoming 
\eqref{metric-general-theta-zero}. Since we are interested in operators with large dimensions, i.e. $m \, R_{\rm AdS} \sim \Delta \gg 1$, we wil be using the geodesic  approximation.

The one-point function of the dual operators can be calculated from a geodesic that  sits at $\theta=\pi/2$ and $\psi=0$ and extends from the boundary of the spacetime until the singularity, that 
is located at $z=r_0^{-1}$. Thus, the one-point function is given by the following expression (see \eqref{NGfin})
\begin{eqnarray} \label{R-1point}
&&\qquad \qquad \qquad \qquad \langle {\cal O}_{\Delta} \rangle \propto e^{ - \Delta \, \ell} =(r_0 \,  \epsilon)^\Delta 
\left(2\, \tilde \epsilon \right)^{\Delta/2} 
\\[5pt]
&&{\rm with}\quad \ell= \int_{\epsilon}^{\frac{1-\tilde{\epsilon}}{r_0}}d z \sqrt{\frac{g_{zz}}{g_{\phi\phi}}} = \int_{\epsilon}^{\frac{1-\tilde{\epsilon}}{r_0}} \frac{d z}{z\, \left(1- r_0^2 \, z^2\right)} = - 
\log r_0 - \log  \epsilon \sqrt{2 \, \tilde{\epsilon}\,} 
\nonumber\, . 
\end{eqnarray}
A couple of important comments are in order.
Notice that in order to obtain a finite result, one has to put besides the usual UV cut-off, represented by 
$\epsilon$, also an IR cut-off. This is implemented through the presence of  $\tilde \epsilon$, since we are not allowed to 
approach close to the singularity because of the logarithmic divergence.
The physical reason for the IR cut-off is that one can not trust the geometry close to the singularity at 
$z= r_0^{-1}$, that is close to the position at which the rotating branes are situated.
The geometry has been generated for a continuous distribution of D3-branes and when one approaches the singularity this approximation breaks down since one can see that the branes are really discrete. 
Notice also that \eqref{R-1point} behaves smoothly in the limit of zero R-charge (i.e. $r_0\rightarrow 0$), namely the one-point vanishes and one gets the $AdS_5$ result for the one point function.
 To express the finite part of the computation is terms of field theory quantities one has to trade the parameter $r_0$ with the chemical potential $\Omega$, as can be seen from the definition of the thermodynamic quantities in \eqref{thermo-quantities}.

The second possibility is to choose a point-like string that sits at  $\theta=0$ and $\psi=0$ inside the five-sphere. In this case the integral that needs to be calculated does not need the presence of the IR cut-off, as before. 
Only the usual UV-cut-off is needed and the result of the calculation reads 
\begin{equation}\label{1-point-R2}
\langle {\cal O}_{\Delta} \rangle \propto e^{ - \Delta \, \ell}=(r_0 \, \epsilon)^\Delta \quad {\rm with} \quad 
\ell= \int_{\epsilon}^{\frac{1}{r_0}}d z \sqrt{\frac{g_{zz}}{g_{\phi\phi}}}= \int_{\epsilon}^{\frac{1}{r_0}} \frac{d z}{z} =- \log \left(r_0 \, \epsilon \right) \, . 
\end{equation}
Notice that \eqref{1-point-R2} behaves smoothly in the limit of zero R-charge (i.e. $r_0\rightarrow 0$), namely the one-point vanishes and one gets the $AdS_5$ result for the one point function.

Notice also that in the case of finite temperature the precise gravity calculation was performed in \cite{Grinberg:2020fdj}. 
In this case for the one point function the presence of a  non-zero a supergravity coupling  of the form $\alpha' \, W^2$ is crucial. 
This term induces the decay of the scalar field into two gravitons. However in our case $W^2$ vanishes
 and another coupling is needed. 
Taking into account that in this background $F_5^2=0$, the first candidate is a correction to the type-IIB action which is schematically of the form $\alpha'^2 \, F_5^4$. It would be nice to perform an explicit computation.
However, notice that such a computation seems extremely difficult since the complete geometry up to the singularity should be determined. 


\subsection{Two-point correlation function}
\label{subsection-R-charge-2point}

In the presence of R-charge (and in the the absence of temperature) there is no distinction 
between temporal and spatial boundary distance, since as can be seen  from 
\eqref{metric-general-theta-pi-over-2} and  \eqref{metric-general-theta-zero} it is the same
function that multiplies $dt^2$ and $d\vec{x}^2$. 
As a result and without loss of generality we restrict the analysis to the case with $\nu =0$, which means that we 
consider operators with only time distance on the boundary. Therefore, the two-point function 
will have only time dependence.  

Similarly to the one-point function calculation, there two cases that we will cover. We will start with the calculation of 
the two-point function described by  a string that sits at $\theta=\pi/2$ and $\psi=0$ and propagates between two points 
of the boundary separated by a time-like distance.
Subsequently,  the point-like string will be placed 
 at $\theta=0$ and $\psi=0$ on the deformed sphere. 

 \subsubsection*{\underline{$\theta=\pi/2$ and $\psi=0$}}

For the two-point correlator described by the string that sits at $\theta=\pi/2$ and $\psi=0$ the expression for $\dot{t}$ in \eqref{tdxd}, after substituting 
the metric components from \eqref{metric_comp_theta_pi_over_2}, becomes
\begin{equation} \label{R-tdot-theta-pi-over2}
\dot t \, = \, \frac{\mu \, z}{\sqrt{h(z)}} 
\quad {\rm with} \quad 
h(z) \, = \, 1- \mu^2 \, z^2 \left(1- r_0^2 \, z^2\right) \ . 
\end{equation}
Integrating \eqref{R-tdot-theta-pi-over2} from the boundary to a generic bulk point  $z$ we get 
\begin{equation} \label{R-t-theta-pi-over2_v1}
t + t_1 = \frac{1}{2 \, r_0} \, \log \Bigg[\frac{2 \,  r_0 \, \sqrt{h(z)}-\mu \left(1-2 \, r_0^2 \, z^2\right)}{2  \, r_0- \mu}\Bigg]
\end{equation}
where we have imposed the boundary condition $t(z=0) = - t_1$. In a similar manner, after substituting 
\eqref{metric_comp_theta_pi_over_2} in \eqref{NGfin} we obtain the following integral expression for the action
\begin{equation}
S \, = \, \Delta \int \frac{dz}{z} \, \frac{1}{ \left(1- r_0^2 \, z^2\right) \, \sqrt{h(z)}} \, .
\end{equation}
Integrating from a point $z=\epsilon$, which lies very close to the boundary, to a generic bulk point  $z$ we 
have 
\begin{equation}  \label{R-S-theta-pi-over2_v1}
\frac{S}{\Delta} = \frac{1}{2} \log 
\left[\frac{1}{\epsilon^{2}} \left(r_0 + \frac{\mu}{2}\right)^{-2} \frac{1+r_0\, \mu  \, z^2 - \sqrt{h(z)}}{1-r_0\, \mu  \, z^2 + \sqrt{h(z)}} \,\,
\frac{\mu +r_0 \big[1-r_0\, \mu  \, z^2 + \sqrt{h(z)}\big]}{\mu -r_0 \big[1+r_0\, \mu  \, z^2 - \sqrt{h(z)}\big]}
\right] \, . 
\end{equation}

Now that we have obtained the equation of motion for $t$ in \eqref{R-t-theta-pi-over2_v1} and determined the exact 
expression for the action functional  in \eqref{R-S-theta-pi-over2_v1}, we are ready to proceed to the calculation of the two-point 
function. We choose to put the operator and its conjugate at the points $(t_1,\vec{0})$ and $(-t_1,\vec{0})$ 
of the boundary, respectively. 
The solution will be described by two geodesics of the form \eqref{R-t-theta-pi-over2_v1} which will touch the boundary 
at the points $t_1$ and $-t_1$, where the two operators are situated. 
By symmetry the two geodesics will meet at the bulk point $(t,z)=(0,z_{\rm max})$, 
where $z_{\rm max}$ will be determined by the the saddle point equation for $z$
\begin{equation} \label{R-meeting-point-theta-pi-over2}
0 = \frac{d S}{d z }= \frac{\partial S}{\partial z}+ \frac{\partial S}{\partial \mu} \, \frac{\partial \mu}{\partial z}
 \quad \Rightarrow \quad 
h(z_{\rm \max})= 0  \quad \Rightarrow \quad 
z_{\rm max}^2 = \frac{\mu -\sqrt{\mu ^2-4 \, r_0^2}}{2 \, \mu  \, r_0^2} \, . 
\end{equation}
The derivative of $\mu$ with respect to $z$ can be read off from \eqref{R-t-theta-pi-over2_v1}, when this is 
evaluated at the joining point. Notice that at the joining point the two geodesics join smoothly, since ${dz/ dt}$ vanishes
(or equivalently  $\dot t  \rightarrow \infty$).

Taking into account that the meeting point for the geodesics is $(t,z)=(0,z_{\rm max})$, we substitute  
\eqref{R-meeting-point-theta-pi-over2} in \eqref{R-t-theta-pi-over2_v1} to obtain an expression for $\tau$ 
as a function of $\mu$, that can be easily inverted
\begin{equation} \label{R-t-theta-pi-over2_v2}
\tau \equiv 2 \, t_1 \,  = \, \frac{1}{2 \, r_0} \, \log\left[\frac{\mu + 2\, r_0 }{\mu - 2\, r_0}\right] 
\quad \Rightarrow \quad 
\mu = 2\, r_0 \, \coth  \tau \, r_0 \, . 
\end{equation}

Substituting \eqref{R-meeting-point-theta-pi-over2} in \eqref{R-S-theta-pi-over2_v1} we obtain an 
expression for the action with respect to $\mu$. Plugging the relation for $\mu$  in terms of $\tau$ from  
\eqref{R-t-theta-pi-over2_v2}, we obtain the two-point correlator in closed form 
\begin{equation} \label{R-2point-theta-pi-over2-v1}
\log \langle{\cal O}_{\Delta}(0) {\cal O}_{\Delta}(\tau)\rangle = -\,  2 \, S_{os} = 2 \, \Delta \, 
\ln \left[\frac{\epsilon}{2}\, \sqrt{\mu^2 - 4 \, r_0^2}\right] = 2 \, \Delta \, 
\ln \left[\frac{\epsilon \, r_0}{\sinh \tau \, r_0}\right] \, . 
\end{equation}
The first three corrections to the $AdS_5$ result for small values of the dimensionless parameter $r_0 \tau$ 
are given by
\begin{equation}  \label{R-2point-theta-pi-over2-v2}
\frac{1}{\Delta}\log \langle{\cal O}_{\Delta}(0) {\cal O}_{\Delta}(\tau)\rangle = - \, 2 \, \log \tau  -\frac{1}{3} \, r_0^2 \, \tau ^2
+\frac{1}{90} \, r_0^4 \, \tau ^4 - \frac{2}{2835} \, r_0^6 \tau ^6 \, . 
\end{equation}
Notice that when the R-charge (chemical potential) is zero, i.e. $r_0=0$, equation \eqref{R-2point-theta-pi-over2-v1} gives 
$ \langle{\cal O}_{\Delta}(0) {\cal O}_{\Delta}(\tau)\rangle= \tau^{-2 \, \Delta}$ 
which is the correct value of the two-point function in the $AdS_5$ background. 
Following the discussion of \cite{Rodriguez-Gomez:2021pfh} we consider a particular large $\Delta$ limit in which 
$r_0 \, \tau \rightarrow 0$ while $\Delta (r_0 \, \tau)^2 $ remains fixed. In this limit the two-point function becomes
\begin{equation}  \label{R-2point-theta-pi-over2-v3}
\langle{\cal O}_{\Delta}(0) {\cal O}_{\Delta}(\tau)\rangle = \frac{1}{\tau^{2 \, \Delta}} \, 
\exp \left[ -\frac{1}{3} \,\Delta \,  r_0^2 \, \tau ^2 \right] \, . 
\end{equation}
In accordance with the discussion around \eqref{1-point}, the operators whose expectation value is non-zero and which  contribute in \eqref{OPE} to give the right hand side of \eqref{R-2point-theta-pi-over2-v3} are schematically of the form $\phi^2, \phi^4, \phi^6, \cdots$, as well as $\phi \, \Box^n \, \phi$ where $\phi$ collectively represent the scalars of the field theory.

 \subsubsection*{\underline{$\theta=0$ and $\psi=0$}}

Lets us now present briefly the results of the calculation for the two point correlator for a string that sits at $\theta=0$ and $\psi=0$.
The corresponding effective metric in this case is \eqref{metric-general-theta-zero}. The expressions for $\dot{t}$ 
and the action, after substituting the metric components from \eqref{metric_comp_theta_zero}, become\footnote{Notice that in order to avoid cluttering  of  notation we are using the same letter for the function that appears in the denominator of $\dot{t}$, i.e. the function $h(z)$. However, in each case we are considering we will declare explicitly the form of this function.}
\begin{equation} \label{R-tdot-action-theta-zero}
\dot t \, = \, \frac{\mu \, z}{\sqrt{\big.h(z)\left(1- r_0^2 \, z^2\right)}}
\quad \& \quad 
\frac{S}{\Delta} = \int \frac{dz}{z} \, \sqrt{\frac{1- r_0^2 \, z^2}{ h(z) }}
\quad {\rm with} \quad h(z) =   1- \left(\mu^2+ r_0^2\right) z^2 \, . 
\end{equation}
The meeting point of the geodesics will be will be determined by setting to zero the function $h(z)$ in the 
denominator of $\dot{t}$, namely 
\begin{equation}  \label{R-meeting-point-theta-zero}
h(z) = 0  \quad \Rightarrow \quad 
z_{\rm max}^2 = \frac{1}{\mu^2 + r_0^2}\, . 
\end{equation}
One could argue that there is an alternative choice for $z_{\rm max}$ (for which $\dot t  \rightarrow \infty$), namely 
$z_{\rm max}=r_0^{-1}$. Notice, however, that this choice does not have a smooth zero $r_0$ limit. In other words, we cannot obtain in a smooth way the  $AdS_5$ limit of the result. Thus, we will not consider this case further.  

Calculating the integrals in \eqref{R-tdot-action-theta-zero} and substituting the expression for the meeting point from 
\eqref{R-meeting-point-theta-zero}, we obtain the following expressions for the time distance of the two operators
\begin{equation}  \label{R-t-theta-zero_v2}
\tau =\frac{2 \, \mu}{r_0 \sqrt{\big. \mu ^2+r_0^2}} \, \log \left[\frac{\mu }{\sqrt{\big. \mu ^2+r_0^2}-r_0}\right]
\end{equation}
and for the action
\begin{equation} \label{R-action-theta-zero_v2}
 - \frac{2 \, S_{os}}{\Delta} = 2 \, \left[\log \left(\frac{\epsilon \, \mu}{2}\right)-
\frac{r_0}{\sqrt{\big.\mu ^2+r_0^2}} \,  \log \left[\frac{\sqrt{\big.\mu ^2+r_0^2}-r_0}{\mu }\right]\right] \, . 
\end{equation}
Contrary to the previous case of  $\theta=\pi/2$ and $\psi=0$, equation \eqref{R-t-theta-zero_v2} cannot be inverted to express $\mu$ as a function of $r_0$ and $\tau$. As a result we will consider the limit of nearly coincident boundary 
points, i.e. the small $\tau$ behavior. Notice that this is equivalent to small $r_0$ or large $\mu$. The result of the expansion reads 
\begin{equation}  \label{R-2point-theta-zero-v2}
\frac{1}{\Delta}\log \langle{\cal O}_{\Delta}(0) {\cal O}_{\Delta}(\tau)\rangle = - \, 2 \, \log \tau  +\frac{1}{6} \, r_0^2 \, \tau ^2
+\frac{1}{90} \, r_0^4 \, \tau ^4 + \frac{47}{22680} \, r_0^6 \tau ^6 \, . 
\end{equation}
Similarly to the previous expansion in  \eqref{R-2point-theta-pi-over2-v2}, equation \eqref{R-2point-theta-zero-v2} has a 
smooth zero R-charge limit for which it reduces to the $AdS_5$ computation. Moreover, the two 
expansions have the same {\it type} of terms (i.e. terms in even powers of the small dimensionless expansion 
parameter $r_0 \tau$)  but with different coefficients.
This is something to be expected since the operators that contribute in the OPE are similar to the ones of the previous result, namely they are schematically of the form $\phi^2, \phi^4, \phi^6, \cdots$  as well as $\phi \, \Box^n \, \phi$. However, they have different R-charges and as a consequence 
different one-point functions.
At the technical level the  two computations differ because they  are performed in two different backgrounds.  

Finally, if we consider the large $\Delta$ limit of \eqref{R-2point-theta-zero-v2}, in which $r_0 \, \tau \rightarrow 0$ and 
$\Delta (r_0 \, \tau)^2 $ remains fixed, we obtain
\begin{equation}   \label{R-2point-theta-zero-v3}
\langle{\cal O}_{\Delta}(0) {\cal O}_{\Delta}(\tau)\rangle = \frac{1}{\tau^{2 \, \Delta}} \, 
\exp \left[ \frac{1}{6} \, \Delta \, r_0^2 \, \tau ^2 \right] \, . 
\end{equation}
It is also a remarkable fact that while the two-point correlators in \eqref{R-2point-theta-zero-v3} increase when their distance is increasing at least for large distances, the ones in \eqref{R-2point-theta-pi-over2-v3} have the opposite behaviour. It would be interesting to comprehend the physical  reasons underlying these behaviours.

\section{Finite temperature}
\label{section:Finite-T}

In this section we focus attention to the case where the background contains only temperature. 
The calculation of the two-point function in this framework has been examined thoroughly in 
\cite{Rodriguez-Gomez:2021pfh} (see also \cite{Rodriguez-Gomez:2021mkk}), 
but here we will discuss some issues concerning the large boundary, temporal and spatial, distance behavior.


\subsection{Two-point correlation function at large temporal distance}
\label{subsection:Finite-T-temporal}

For reasons of completeness, in the beginning of this subsection we will review part of the material that was presented 
in \cite{Rodriguez-Gomez:2021pfh} and concerns the two-point function calculation. Notice that when only temperature 
is present, there is no distinction between the two orientations of the string, i.e. $(\theta=\pi/2\, , \, \psi=0)$ and 
$(\theta=0 \, , \,  \psi=0)$. They are completely equivalent. 

For temporal boundary distance the integrals in \eqref{tdxd}
and \eqref{NGfin} can be computed analytically. The integral of the action reads
 \begin{equation}
 \label{T-action-integral-mu}
- \, \frac{2 \, S}{\Delta} = \log \left[\epsilon ^2 \, 
\frac{2 -\mu ^2\, z^2 +2  \sqrt{h(z)}}{4 \, z^2}\right]
\quad {\rm with} \quad h(z) = 1 - \eta^4 \, z^4 - \mu ^2 \, \,z^2 
\end{equation}
 while for the integral of $\dot t$ the result is 
 \begin{eqnarray}  \label{T-t-t1-mu}
 t-t_1 &=& - \, \frac{1}{4 \, \eta} \, \log \left[\frac{\mu ^2-  \left(\mu ^2-2 \, \eta^2\right) \eta ^2 \, z^2 + 2 \, \eta \,  \mu  
 \sqrt{h(z)} + 2 \,  \eta ^2}{  \left(1+ \eta ^2 \, z^2 \right) 
 \left(\mu ^2 + 2 \, \eta \,  \mu + 2 \, \eta^2\right)}\right] 
  \nonumber  \\[7pt]
   && + \,  \frac{i}{4 \, \eta} \, \log \left[\frac{\mu ^2 + \left(\mu ^2 + 2 \, \eta^2\right) \eta ^2 \, z^2 +  2\, i  \, \eta \,  \mu  
 \sqrt{h(z)} - 2 \,  \eta ^2}{  \left(1- \eta ^2 \, z^2 \right) 
 \left(\mu ^2 +2 \, i \, \eta \,  \mu - 2 \, \eta^2\right)}\right] \, . 
\end{eqnarray}
Following the reasoning of subsection \ref{subsection-R-charge-2point}, the two operators will be placed at the boundary points 
$(-t_1,\vec{0})$ and $(t_1,\vec{0})$ and the meeting point in the bulk will have coordinates $(0,z_{\rm max})$. 
The value of $z_{\rm max}$ will be determined from setting to zero the value of the function $h(z)$, as it is defined 
in \eqref{T-action-integral-mu}. The result of this computation reads
\begin{equation} \label{T-2point-meeting-point-tau}
h(z)\, = \, 0 \quad \Rightarrow \quad  
z_{\max}^2 =  \frac{1}{2 \, \eta^4} \, \left[- \mu^2 + \sqrt{\big. \mu^4 + 4 \, \eta^4}\right] \, . 
\end{equation}
Substituting \eqref{T-2point-meeting-point-tau} into \eqref{T-t-t1-mu}, we obtain the following expression for the temporal boundary distance $\tau$
\begin{equation}  \label{T-tau-mu}
\tau= 2 \, t_1 = \frac{1}{2 \, \eta} \, \Bigg[  i \,  
\log \left[\frac{\sqrt{4 \, \eta ^4+\mu ^4}}{\mu ^2-2 \, \eta ^2+2 \, i \, \eta \, \mu}\right] 
- \log \left[\frac{\mu ^2+ 2 \, \eta ^2-2 \, \eta \,  \mu}{\sqrt{4 \, \eta ^4+\mu ^4}}\right]\Bigg]
\end{equation}
while substituting \eqref{T-2point-meeting-point-tau} into \eqref{T-action-integral-mu} we obtain the 
following expression for the on-shell action
\begin{equation} \label{T-action-saddle-v1}
\frac{S_{os}}{\Delta} =  - \, \log \left[\frac{\epsilon}{2} \, \Big[4 \, \eta ^4+\mu ^4 \Big]^{\frac{1}{4}} \right]\, .
\end{equation}
Contrary to the case of finite R-charge that we presented in subsection \ref{subsection-R-charge-2point}, 
we cannot invert \eqref{T-tau-mu} and obtain an expression of $\mu$ in terms of 
$\tau$. However we can expand both \eqref{T-tau-mu} and \eqref{T-action-saddle-v1} for small and large values of $\eta$
(i.e. low and high temperature respectively) to obtain perturbative expressions. 

Expanding \eqref{T-tau-mu} and \eqref{T-action-saddle-v1} for small values of $\eta$ (or equivalently large values of $\mu$), 
we obtain the following expression for the perturbative expansion of the two-point function at small temperature
\begin{equation}  \label{T-2point-perturbative-small-eta}
\frac{1}{\Delta}\log \langle{\cal O}_{\Delta}(0) {\cal O}_{\Delta}(\tau)\rangle = - \, 2 \, \log \tau  + 
\frac{1}{40} \, \pi^4 \,  T^4 \, \tau^4 \Bigg[1+\frac{11}{360}\, \pi^4 \,  T^4 \, \tau^4 
+\frac{89}{46800}\, \pi^8 \,  T^8 \, \tau^8 \Bigg]
\end{equation}
where we have used the relation between $\eta$ and $T$, namely $\eta = \pi \, T$. The results we have discussed  
so far in the current subsection were first presented in \cite{Rodriguez-Gomez:2021pfh}. 

Expanding \eqref{T-tau-mu} for large values of $\eta$ we have
\begin{equation}  \label{T-tau-perturbative-large-eta}
\delta \tau = \frac{\pi}{2} - \tau \, \eta= \frac{\pi}{2} \left(1 - 2 \, \tau \, T\right) = \frac{1}{6} \, \frac{\mu ^3}{ \eta^3} -
 \frac{1}{56} \, \frac{\mu ^7}{ \eta^7}  +  \frac{1}{352} \, \frac{\mu^{11}}{ \eta^{11}}  \, + 
 {\cal O} \left(\frac{\mu ^{15}}{ \eta^{15}} \right)
\end{equation}
and the two-point function reads in this case
\begin{equation}  \label{T-tau-2point-perturbative-large-eta}
\frac{\langle {\cal O}(0) {\cal O}(\tau)\rangle}{T^{2 \, \Delta}} \, \frac{2^{\Delta}}{\pi^{2 \, \Delta}} =
1+ \frac{3^{4/3}}{2^{5/3}}\, \Delta \, \delta \tau^{4/3} + \frac{3^{8/3}}{2^{10/3}}\, \Delta\, \left(\frac{1}{7} + \frac{\Delta}{2}\right) \, \delta \tau^{8/3} +  {\cal O} \left(\delta \tau^{8/4} \right) \, . 
\end{equation}
Notice that in \eqref{T-tau-2point-perturbative-large-eta} the ``small" parameter in the expansion of the two-point function is 
not $\tau\, T$ but the difference of  $\tau\, T$ from $1/2$, which due to the periodicity of the time coordinate 
is the maximum value that the dimensionless quantity $\tau \, T$ can take. More precisely, $\frac{1}{2 T}$ is the maximum temporal distance of the two operators and corresponds to half of the time period.
Another observation that emerges from \eqref{T-tau-2point-perturbative-large-eta} is that the two-point function 
in the limit of high temperature seems to be approximated by two one-point functions. 
However one should be extremely careful at this point because the one-point function in a thermal background 
has a prefactor that contains $\alpha'$, due to the presence of the square of the Weyl tensor (see the discussion in 
\cite{Grinberg:2020fdj, Myers:2016wsu}). The limit in \eqref{T-tau-2point-perturbative-large-eta} only 
reproduces the exponential of the action, not the prefactor.


\subsection{Two-point correlation function at large spatial distance}
\label{subsection:Finite-T-spatial}

In this section, we consider the computation of the holographic two-point function for a set of operators 
that their boundary distance is spatial, i.e. we will take $\mu =0$. 
We will draw interesting conclusions comparing the two-point functions with spatial and temporal 
dependence. In this case the integral of the action reads
\begin{equation}
 \label{T-action-integral-nu}
- \frac{S}{\Delta}  = 
\frac{\eta}{\sqrt{\big.\eta^2 +\nu^2}} \left[{\mathbf \Pi} 
\left[2; a_1 \Big| b \right]  - {\mathbf \Pi} \left[2; a_2 \Big| b \right]\right]
\end{equation}
while the integral of $\dot x$ becomes 
\begin{equation} \label{T-x-x1-nu}
x - x_1 \, = \,  
\frac{ i \, \nu}{\eta  \sqrt{\eta^2+\nu ^2}} \, \Big[{\mathbf F} \left(\left.a_3\right|1-b\right)- 
{\mathbf F} \left(\left.a_4\right|1-b\right)\Big]
\end{equation}
with the constants taking the following values
\begin{eqnarray}
&& a_1 = \arcsin \left[\sqrt{\frac{1 + \eta ^2 \, z^2 }{2}} \right]   \quad 
a_2 = \arcsin \left[\sqrt{\frac{1 + \eta ^2 \, \epsilon^2}{2}} \right]\, , \quad
a_4 = \arcsin \left[\sqrt{\frac{2}{2-b}}  \right]  
\nonumber\\[5pt]
&&
a_3 = \arcsin \left[\sqrt{\frac{2}{2-b}} \, \frac{1}{\sqrt{1- \nu^2 \, z^2}}\right]   \quad \& \quad  
b = \frac{2 \, \nu^2}{\eta^2+\nu^2} \, . 
\end{eqnarray}
By using the fact that the maximum of the arc is situated at $(z,x)=(\nu^{-1},0)$ one obtains from \eqref{T-x-x1-nu} that half of the distance between the operators is 
\begin{equation} \label{x1}
x_1=- \, \frac{i \, \nu }{\eta  \, \sqrt{\eta ^2+\nu ^2}} \Bigg[{\mathbf F} \left(\arcsin\left(\frac{\sqrt{\eta ^2+\nu ^2}}{\eta }\right)\Bigg|\frac{\eta^2-\nu ^2}{\eta^2+\nu^2}\right)-i \, {\mathbf K} \left(\frac{2 \, \nu ^2}{\eta ^2+\nu ^2}\right)\Bigg] \, . 
\end{equation}
The last equation implies that $x_1<0$ which is consistent with the fact that $\nu>0$.

In what follows, we will be interested in the case where the two operators are put at a large spatial distance 
$\delta x=2 x_1$. This can be achieved only if the integration constant $\nu$ is very close to 
$\eta$. We, thus, take $\nu=\eta+\varepsilon$ with $\varepsilon$ being infinitesimally small and positive, $\varepsilon> 0$. By inserting this value for $\nu$ in \eqref{x1} and expanding the resulting expression in $\varepsilon$ one obtains by keeping the leading in $\varepsilon$ terms
\begin{equation} \label{eps1}
x_1 \, = \, \frac{1}{2 \, \sqrt{2} \, \eta} \, \log\frac{\varepsilon}{\eta} 
\quad \Rightarrow \quad 
\varepsilon \, = \, \eta \,  e^{-\sqrt{2} \, \eta \, |\delta x|}\, .
\end{equation}
In order to determine the two-point function one should expand the incomplete elliptic integrals of the third kind 
appearing in \eqref{T-action-integral-nu} for small values of $\varepsilon$. 
To this end, we use the following identity which gives the expansion around $m=1$
\begin{eqnarray}\label{id}
&&\mathbf \Pi[n, z|m] \approx\frac{1}{2 (n - 1)} \left[\sqrt{n} \, \log{\frac{1 + \sqrt{n} \sin{z}}{1 - \sqrt{n} \sin{z}}} - 
2 \log{(\sec{z} + \tan{z})}\right] + 
\nonumber \\[5pt]
&& \quad \qquad  \quad + \,  \frac{1}{6} \, \sin{z}^3 \, {\rm F}_1\left[\frac{3}{2}, 2, 1, \frac{5}{2}, \sin{z}^2, n \sin{z}^2\right] 
\left(m - 1\right) + 
\nonumber  \\[5pt]
&&
\quad \qquad  \quad + \,  \frac{3}{40} \, \sin{z}^5 \, {\rm F}_1\left [\frac{5}{2}, 3, 1, \frac{7}{2}, \sin{z}^2, n \sin{z}^2\right]  \left(m - 1\right)^2 + 
\cdots 
\end{eqnarray}
where ${\rm F}_1$ is the Appell hypergeometric function of two variables.
Notice that in our case $m=1+\frac{\varepsilon}{\eta}+\cdots$ and that the range of validity of \eqref{id} is when 
$\rm |Re \, z|\le\frac{\pi}{2}$, which is the case for both the elliptic integrals appearing in \eqref{T-action-integral-nu}.
In fact only the first line of \eqref{id} contribute to the leading term of the two-point function, thus we obtain
\begin{equation}
 \label{S-expa}
- \, \frac{2 \, S}{\Delta} = \log \left(1-\sqrt{2}\sin{\alpha_2}\right)  -\sqrt{2} \, \log 
\left(\sec{z_1} + \tan{z_1}\right) \, .
\end{equation}
Notice that the first term in the right hand side of \eqref{S-expa} originates from the first term in the parenthesis appearing in the first line of \eqref{id}, while the second term originates from the second term in the first line of \eqref{id}.
Expanding \eqref{S-expa} for small $\varepsilon$ and $\epsilon$ and using \eqref{eps1} we obtain the following expression for the two-point function
\begin{equation}
 \label{G2}
G_2 \propto e^{-2 \, S}= \Big[-\frac{\epsilon^2 \eta^2}{2 \sqrt{2}}\Big] ^{\Delta}e^{- \, \Delta \, \eta \, |\delta x|} \, . 
\end{equation}
From \eqref{G2} we see that the two point function falls exponentially with the distance between the two operators for large spatial distances.

A last comment is in order. Notice that due to the exponential suppression of the two-point correlator there is a critical value of the distance $|\delta x|$ for which the two-point function is dominated by a different saddle point. Namely, there is another saddle point which is comprised by two straight geodesics having $\nu=0$. Each of those starts from the point where the operators are situated and propagate to cross the horizon of the black hole. In this case the two-point function behaves as $G_2\sim \langle \mathcal O\rangle \langle \mathcal {\bar O}\rangle+c \, e^{-m_{th} |\delta x|}$, where $\langle \mathcal O\rangle$ is the one-point function of the operators and $m_{th}$ is the thermal mass. It is evident that despite the fact that each of the one-point functions is proportional to $\alpha'$ this second saddle point dominates beyond a critical value of the distance between the two operators. This happens because, as mentioned above, the connected saddle point fall exponentially with the distance while the disconnected one is essentially independent of it.

\section{Finite temperature and finite R-charge chemical potential}
\label{section:R-charge-T} 

The aim of this section is to compute one-point and two-point correlation functions of scalar operators at strong coupling in the presence of both finite temperature and finite density (chemical potential) and specify the effects of the latter to the aforementioned correlation functions.
The gravity computations will be in terms of the parameters $\eta$ and $r_0$ appearing in the supergravity solution. Subsequently,  we will use the expressions in section \ref{section:backgrounds} to express them as functions of the thermodynamic quantities $T$ and $J$ or $T$ and 
$\Omega$. 

\subsection{One-point correlation function}

As usual there are two cases to cover.

 \subsubsection*{\underline{$\theta=\pi/2$ and $\psi=0$}}

When the dual to the operator string sits at $\theta=\pi/2$ and $\psi = 0$ the proper length between the boundary and the singularity at $r_0^{-1}$, that appears in the exponential of \eqref{R-1point},  reads
\begin{equation} \label{RT-1point_v1}
\ell = \int^{\frac{1-\tilde{\epsilon}}{r_0}}_{\epsilon} \frac{d z}{z\, \left(1- r_0^2 \, z^2\right)\sqrt{\big. 1- \eta^4 \, z^4}} \, . 
\end{equation}
Notice that similar to the case in which only the R-charge is present, there is the need of both an IR and a UV cut-off.  
The calculation of  $\ell$ can break in two pieces: the first one is the integral from the boundary of the space until 
the horizon of the black hole, that acquires a real value, and the second piece is the integral from the horizon until the singularity, that acquires an imaginary value. In fact, the result of this second piece is related to the 
proper time of a particle traveling from the horizon to the singularity \cite{Grinberg:2020fdj}. 

The integral in \eqref{RT-1point_v1} can be calculated analytically but since it has a very complicated expression for generic values of the parameters $r_0$ and $\eta$, we will present an expansion in small values of the dimensionless 
parameter $\delta = r_0/\eta$.  The real part reads
\begin{equation} \label{RT-1point-real_v1}
\ell_R =  -\log \left(\frac{\eta  \, \epsilon }{\sqrt{2}}\right)
+ \frac{\pi}{4} \, \delta^2
+ \frac{1}{2} \, \delta^4
+ \frac{\pi}{8} \, \delta^6
+ \frac{1}{3} \, \delta^8
+ \frac{3\,\pi}{32} \, \delta^{10}
+ \frac{4}{15} \, \delta^{12}
\end{equation}
while for the imaginary part the result of the calculation is
\begin{equation}  \label{RT-1point-imaginary_v1}
\ell_I = - \frac{i \, \pi }{4}\Bigg[1- 
\frac{2\, \delta ^2}{\pi} \left(1+ 2 \log \delta \right)
- \frac{\delta ^6}{3\, \pi } \left(7+ 6 \log \delta \right)
- \frac{\delta ^{10}}{20\, \pi } \left(43+ 30 \log \delta \right)  \Bigg] + 
\frac{i\, \delta^2 \, \log \tilde{\epsilon}}{2 \, \sqrt{1-\delta^4}} \, . 
\end{equation}
The last term depends on the IR cut-off $\tilde \epsilon$ and its origin is explained below \eqref{R-1point}.
Notice that in the limit $\delta \rightarrow 0$, with the number of colours being large and fixed, the aforementioned term goes to zero and one recovers the result for the black hole \cite{Grinberg:2020fdj}. 

Exponentiating  \eqref{RT-1point-real_v1} and using \eqref{thermo-quantities} to express $\eta$ as a function of 
the temperature $T$ and $r_0$ as a function of the chemical potential $\Omega$, we obtain the following expression for the real contribution to the one-point function
\begin{equation} \label{RT-1point-real_v2}
\langle {\cal O}_{\Delta} \rangle_{R} \propto  \frac{\pi^{\Delta}\, \epsilon^{\Delta}}{2^{\Delta \over 2}} T^{\Delta} \Bigg[1+\frac{2-\pi}{4 \, \pi ^2} \frac{\Delta \, \Omega^2}{T^2}+\frac{ (\pi -2)^2 \Delta +8 (\pi -3)}{32 \, \pi ^4}  \frac{\Delta \, \Omega^4}{T^4} + {\cal O} \left(\frac{\Omega^6}{T^6} \right)\Bigg] \, . 
\end{equation}
Notice that the above result reduces to the one-point function calculated in \cite{Grinberg:2020fdj} in the limit $\delta \rightarrow 0$. The leading term of \eqref{RT-1point-real_v2} has the correct scaling $T^\Delta$ for an operator of dimension $\Delta$.  
The remaining terms in the bracket of \eqref{RT-1point-real_v2} represent the effect of turning on a non-zero chemical potential.
Notice also that in the expression above we have expressed the one-point function in terms of thermodynamic quantities of the Grand Canonical Ensemble. 
Similarly, from the relation $\ell=\ell_R-i \tau_s$ one easily deduces that 
\begin{equation}
\tau_s=  \frac{ \, \pi }{4}\Bigg[1- 
\frac{2\, \delta ^2}{\pi} \left(1+ 2 \log \delta \right)
- \frac{\delta ^6}{3\, \pi } \left(7+ 6 \log \delta \right)
- \frac{\delta ^{10}}{20\, \pi } \left(43+ 30 \log \delta \right)  \Bigg] - 
\frac{\delta^2 \, \log \tilde{\epsilon}}{2 \, \sqrt{1-\delta^4}} \, . 
\end{equation}
Again the leading term in this expression represents the travelling time from the horizon of the black hole to its singularity and agrees with the result of \cite{Grinberg:2020fdj} while the remaining terms encode the effect of the finite density (chemical potential) to this time.
We conclude that as in the case of finite temperature \cite{Grinberg:2020fdj} the one-point function of scalar operators  provides information for processes behind the horizon of the black hole.

 \subsubsection*{\underline{$\theta=0$ and $\psi=0$}}

When the dual to the operator string sits at $\theta=0$ and $\psi = 0$ the calculation of the proper length is more intricate. It is easy to check 
that while it is consistent with the equations of motion in \eqref{tdxd} to set $\nu=0$, it is inconsistent to set $\mu=0$ because $g_{t\phi}$ in this case is different from zero. In particular this implies that there is no particle, and as a result no geodesic, that can travel in a {\it straight} line from the point of the boundary where the dual field theory operator is situated down to the singularity since $\dot t$ is generically different from zero.  In order for the one-point function to be properly defined, we fix the value of $\mu$ in such a way that $\dot t$ goes to infinity exactly at the horizon. This happens for $\mu = r_0$. Using these values for $\nu$ and $\mu$ the proper length that appears in the exponential of \eqref{R-1point},  reads
\begin{equation}  \label{RT-1point_v2}
\ell = \int^{\frac{1}{\eta}}_{\epsilon} \frac{\left[1 + \left(\eta^2 - r_0^2 \right) z^2\right]}{
\sqrt{\big.\left(1- r_0^2 \, z^2\right) \left( 1- \eta^2 \, z^2\right)\left[1 + \left(\eta^2 - 2\, r_0^2 \right) z^2\right]}} \frac{dz}{z \left(1+\eta^2 \, z^2\right)} \, . 
\end{equation}
As previously, the integral in  \eqref{RT-1point_v2} can be calculated analytically for generic values of $r_0$ and $\eta$, 
but we will present its expansions for small values of the dimensionless parameter $\delta$. The result reads
\begin{equation} \label{RT-1point-real_v2a}
\ell_R =  -\log \left(\frac{\eta  \, \epsilon }{\sqrt{2}}\right)+ \frac{\pi}{8}\, \delta ^2+
\frac{13}{48}  \delta^4 +
\left(\frac{13 \pi }{128}-\frac{1}{10}\right) \delta ^6 +
\left(\frac{1217}{13440}+\frac{\pi }{32}\right) \delta ^8 
\end{equation}
As a result, the contribution to the one-point function in terms of the thermodynamic quantities $T$ and $\Omega$ is given by
\begin{equation} \label{RT-1point-real_v2b}
\langle {\cal O}_{\Delta} \rangle_{R} \propto  \frac{\pi^{\Delta}\, \epsilon^{\Delta}}{2^{\Delta \over 2}} T^{\Delta} \Bigg[1+\frac{4-\pi}{8 \, \pi ^2} \frac{\Delta \, \Omega^2}{T^2}+\frac{ 3 (\pi -4)^2 \Delta +8 (6\, \pi -25)}{384 \, \pi ^4}  \frac{\Delta \, \Omega^4}{T^4} + {\cal O} \left(\frac{\Omega^6}{T^6} \right)\Bigg] \, . 
\end{equation}

Notice that the geodesic describing this one-point function terminates at the horizon. It does not  cross the horizon and thus it never falls into the singularity. Consequently, this kind of one-point function can not give us information about the traveling time from the horizon to the singularity.

However, one may blindly analytically  continue and calculate the integral \eqref{RT-1point_v2} from the horizon at $z=\eta^{-1}$ to the singularity at $z=r_0^{-1}$.
It is a peculiar fact that the result is found to be
\begin{equation}  \label{RT-1point-imaginary_v2}
\ell_I = - \frac{i \, \pi }{4}\Bigg[1 
- \left(1+ \log \left(\frac{\delta^2}{8}\right)\right) \frac{\delta^2}{\pi}+ \frac{2 \, \delta ^4}{3 \, \pi } -
\left(226+ 195 \log \left(\frac{\delta^2}{8}\right)\right)\frac{\delta^6}{240 \, \pi} \Bigg]
\end{equation}
whose leading term is related to the correct proper time from the horizon to the singularity of a black hole  \cite{Grinberg:2020fdj} while the  
remaining terms represent the finite density corrections. It would be nice to understand if this result is of any physical relevance.

\subsection{Two-point correlation function}

In the presence of both temperature and R-charge there is a distinction between the temporal and the spatial boundary distance, since the functions that multiply $dt^2$ and $d\vec{x}^2$  in the metric are different. This happens for both orientations of the string, as can be seen from \eqref{metric-general-theta-pi-over-2} and \eqref{metric-general-theta-zero}. 
We will first present the case with only temporal boundary distance and in the following we will include  spatial dependence. This last step is important in order to make contact with the OPE expansion and the connection with field theory, that is presented in section \ref{section:OPE}.

\subsubsection{Temporal boundary distance}
\label{subsubsection:Finite-T-R-temporal}

As usual we will consider two cases. 

 \subsubsection*{\underline{$\theta=\pi/2$ and $\psi=0$}}

We will start the analysis with the case in which the string used to evaluate the two-point correlator has the orientation $\theta=\pi/2$ and $\psi=0$. The expression for 
$\dot t$, after substituting the relevant metric components from \eqref{metric-general-theta-pi-over-2}, becomes 
\begin{equation} \label{RT-tdot-theta-pi-over2_v1}
\dot t \, = \, \frac{\mu \, z}{\left(1 - \eta^4 \, z^4\right)\sqrt{\big. h(z) }} 
\quad {\rm with} \quad  h(z) = 1 - \eta^4 \, z^4 - \mu^2 \, z^2 \left(1- r_0^2 \, z^2\right) \, . 
\end{equation}
Lets us remind the reader that the function $h(z)$ has a different form in each case that we consider. 
The integral expression for the action becomes
\begin{equation}  \label{RT-S-theta-pi-over2_v1}
S \, = \, \Delta \int \frac{dz}{z} \, \frac{1}{ \left(1- r_0^2 \, z^2\right) \, \sqrt{\big. h(z)}} \, . 
\end{equation}
The turning point of the string in the bulk is given by the point in which the function $h(z)$ vanishes, namely 
 \begin{equation}  \label{RT-meeting-point-theta-pi-over2}
h(z)= 0  \quad \Rightarrow \quad 
z_{\rm max}^2 = \frac{1}{2 \left( \mu^2  \, r_0^2 - \eta^4 \right)} 
\left[\mu^2 -\sqrt{\big.\mu^4-4 \, r_0^2 \, \mu^2 + 4 \, \eta^4} \right] \, . 
\end{equation}
Integrating \eqref{RT-tdot-theta-pi-over2_v1} from the boundary until the meeting point  \eqref{RT-meeting-point-theta-pi-over2}, 
we obtain the following expression for the temporal boundary distance
\begin{equation}  \label{RT-t-theta-pi-over2_v1}
4 \, \tau = \frac{1}{\sqrt{\big. r_0^2 + \eta^2}} \ln \left[\frac{2 \, \eta^2+ \mu^2 +2 \,  \sqrt{\big. r_0^2 + \eta^2}}{2 \, \eta^2+ \mu^2 -2 \,  \sqrt{\big. r_0^2 + \eta^2}}\right] +
\frac{1}{\sqrt{\big. r_0^2 - \eta^2}} \ln \left[\frac{2 \, \eta^2- \mu^2 -2 \,  \sqrt{\big. r_0^2 - \eta^2}}{2 \, \eta^2- \mu^2 +2 \,  \sqrt{\big. r_0^2 - \eta^2}}\right]
\end{equation}
while for the integral of \eqref{RT-S-theta-pi-over2_v1} we have
\begin{equation}  \label{RT-S-theta-pi-over2_v2}
 \frac{2 \, S_{os}}{\Delta} =
\ln \left[\frac{4}{\epsilon^2 \, \sqrt{\big.\mu^4-4 \, r_0^2 \, \mu^2 + 4 \, \eta^4}}\right]  + \frac{i \, r_0^2}{\sqrt{\big. \eta^4 - r_0^4}} \, 
\ln \left[\frac{\mu^2 - 2 \, r_0^2 - 2 \, i \, \sqrt{\big. \eta^4 - r_0^4}}{\sqrt{\big.\mu^4-4 \, r_0^2 \, \mu^2 + 4 \, \eta^4}}\right] \, . 
\end{equation}
Finally, we combine \eqref{RT-t-theta-pi-over2_v1} and \eqref{RT-S-theta-pi-over2_v2} to express the 
two-point function as a series expansion around the $AdS_5$ result for operators that are close to each other, 
i.e. small $\tau$ (this is equivalent with an expansion for small $r_0$ and $\eta$)
\begin{equation}   \label{RT-2point-theta-pi-over2_v1}
\frac{1}{\Delta}\log \langle{\cal O}_{\Delta}(0) {\cal O}_{\Delta}(\tau)\rangle = - \, 2 \, \log \tau  -\frac{1}{3} \, r_0^2 \, \tau ^2
+\frac{1}{90} \, r_0^4 \, \tau ^4 + \frac{1}{40} \, \eta^4 \, \tau^4- \frac{2}{2835} \, r_0^6 \tau ^6 - \frac{1}{105} \, r_0^2 \, \eta^4 \tau^6 \, . 
\end{equation}
Notice that  setting to zero the parameter $\eta$ in  \eqref{RT-2point-theta-pi-over2_v1} we obtain 
\eqref{R-2point-theta-pi-over2-v2}, while setting to zero the parameter $r_0$ we obtain \eqref{T-2point-perturbative-small-eta}.
The two-point function in  \eqref{RT-2point-theta-pi-over2_v1} can be rewritten in terms of the thermodynamic quantities 
$T$ and $\Omega$. The result reads
\begin{eqnarray}   \label{RT-2point-theta-pi-over2_v2}
\frac{1}{\Delta}\log \langle{\cal O}_{\Delta}(0) {\cal O}_{\Delta}(\tau)\rangle &=& - \, 2 \, \log \tau  -\frac{1}{3} \, \Omega^2 \, \tau ^2
+\frac{1}{360} \left(9 \, \pi ^4 \, T^4+18 \, \pi ^2 \, T^2 \, \Omega ^2+13 \, \Omega ^4\right) \tau ^4 
\nonumber \\[5pt] 
&&- \,\frac{1}{2835} \, \Omega^2 \, \left(27 \, \pi ^4 \, T^4+54 \, \pi ^2 \, T^2 \, \Omega ^2+29 \, \Omega ^4\right)  \tau^6 \, . 
\end{eqnarray}
Finally we consider the following large $\Delta$ limit 
\begin{equation} \label{largeD_limit}
\Delta \rightarrow \infty\, , \quad \Omega \, \tau \rightarrow 0 \, , \quad \Delta (\Omega \, \tau)^2 = {\rm fixed} \, , \quad
T \, \tau \rightarrow 0 \, , \quad \Delta (T \, \tau)^4 = {\rm fixed}
\end{equation}
with which it is possible to exponentiate \eqref{RT-2point-theta-pi-over2_v2}
\begin{equation}   \label{RT-2point-theta-zero-v3}
\langle{\cal O}_{\Delta}(0) {\cal O}_{\Delta}(\tau)\rangle = \frac{1}{\tau^{2 \, \Delta}} \, 
\exp \left[- \frac{1}{3} \, \Delta \, \Omega^2 \, \tau ^2 + \frac{\pi^4}{40} \,\Delta \, T^4 \, \tau^4 \right] \, . 
\end{equation}

 \subsubsection*{\underline{$\theta=0$ and $\psi=0$}}

Now we will move to the second choice for the string orientation, namely $\theta=0$ and $\psi=0$. 
The expression for $\dot t$ becomes
 \begin{equation} \label{RT-tdot-theta-pi-zero_v1}
\dot t \, = \, \frac{\mu \, z - r_0 \, \eta^2 \, z^3}{\left(1 - \eta^4 \, z^4\right)\sqrt{\big. \left(1 - r_0^2 \, z^2\right)h(z) }} 
\quad {\rm with} \quad  h(z) = 1 -\eta^4 \, z^4 - \left(\mu^2 + r_0^2 \right) z^2 +2 \, r_0 \, \eta^2 \, \mu \, z^4
\end{equation}
while for the integral of the action we have
\begin{equation} \label{RT-S-theta-zero_v1}
S \, = \, \Delta \int \frac{dz}{z} \,\frac{1 - r_0^2 \, z^2 - \eta^4 \, z^4 +r_0 \, \eta^2 \, \mu \, z^4}{\left(1 - \eta^4 \, z^4\right)\sqrt{\big. \left(1 - r_0^2 \, z^2\right)h(z) }} \, . 
\end{equation}
The vanishing of the function $h(z)$ is going to give the $z$ coordinate of the meeting point
\begin{equation}  \label{RT-meeting-point-theta-zero}
h(z)= 0  \quad \Rightarrow \quad 
z_{\rm max}^2 = - \, \frac{\mu ^2+r_0^2 -
\sqrt{\big.4 \, \eta ^4+\mu ^4-8 \, \eta ^2 \, \mu  \, r_0+2 \, \mu ^2 \, r_0^2 + r_0^4}}{2 \, \eta ^4-4 \, \eta ^2 \, \mu  \, r_0} \, . 
\end{equation}
It is easy to check that expanding  \eqref{RT-meeting-point-theta-zero} for small values of $\eta$, one obtains the 
$z$ coordinate of the meeting point that we calculated in \eqref{R-meeting-point-theta-zero}. The integrals in 
\eqref{RT-tdot-theta-pi-zero_v1} and  \eqref{RT-S-theta-zero_v1} cannot be calculated analytically and for this 
reason we calculate them in the limit of nearly coincident operators, i.e. $\tau \rightarrow 0$ or equivalently 
$\mu \rightarrow \infty$. The result for the temporal distance is 
\begin{eqnarray}  \label{RT-t-theta-zero_v1}
&& \frac{\mu \, \tau}{2} = 1-\frac{2 \, r_0^2}{3 \, \mu ^2}+\frac{2 \, \eta ^2 \, r_0}{\mu ^3}+
\frac{4 \left(2 \, r_0^4-3 \, \eta ^4\right)}{15 \, \mu ^4}-
\frac{16 \left(\eta ^2 \, r_0^3\right)}{3 \, \mu ^5}-
\frac{16 \left(r_0^6-25 \, \eta ^4 \, r_0^2\right)}{35 \, \mu^6}
\nonumber \\[5pt]
&& 
\qquad \qquad
+\frac{48 \, \eta ^2 \, r_0^5-40 \,  \eta ^6 \, r_0}{5\, \mu ^7}+
\frac{16 \left(35 \, \eta ^8-924 \, \eta ^4 \, r_0^4+8 \, r_0^8\right)}{315 \, \mu ^8}
\end{eqnarray}
while the on-shell action becomes
\begin{equation}  \label{RT-S-theta-zero_v2}
- \frac{2 \, S_{os}}{\Delta} = 
2 \Bigg[\log \left(\frac{\epsilon \, \mu}{2} \right)+
\frac{r_0^2}{\mu ^2} - 
\frac{8 \, \eta ^2 \, r_0}{3 \, \mu ^3} +
\frac{3 \, \eta ^4-2 \, r_0^4}{3 \, \mu ^4}+ 
\frac{32 \, \eta ^2 \, r_0^3}{5 \, \mu ^5} - 
\frac{8 \left(25 \, \eta ^4 \, r_0^2- r_0^6\right)}{15 \, \mu ^6} \Bigg]
\end{equation}
As previously, we combine \eqref{RT-t-theta-zero_v1} and \eqref{RT-S-theta-zero_v2} to calculate the 
two-point function as a series expansion for small values of $\tau$, which is equivalent with an expansion for 
small values of $r_0$ and $\eta$
\begin{eqnarray}   \label{RT-2point-theta-zero_v1}
&& \frac{1}{\Delta}\log \langle{\cal O}_{\Delta}(0) {\cal O}_{\Delta}(\tau)\rangle = - \, 2 \, \log \tau  + 
\frac{1}{6} \, r_0^2 \, \tau ^2 -
\frac{1}{6} \, r_0 \, \eta^2 \, \tau ^3  +
\frac{1}{90} \, r_0^4 \, \tau ^4 + 
\frac{1}{40} \, \eta^4 \, \tau^4 
\nonumber \\[5pt]
&& 
\qquad \qquad 
-  \frac{1}{60} \, r_0^3 \, \eta^2 \, \tau^5 + 
\frac{47}{22680} \, r_0^6 \, \tau^6 + 
\frac{11}{560} \, r_0^2 \, \eta^4 \, \tau^6 \, . 
\end{eqnarray}
Notice that the two-point function in  \eqref{RT-2point-theta-zero_v1} contains odd powers of $\tau$ in the expansion, 
besides the even powers that they also exist in  \eqref{RT-2point-theta-pi-over2_v1}. 
This is related to presence of the non-diagonal $g_{t\phi}$ term in \eqref{metric-general-theta-zero}.  
Setting to zero the parameter $\eta$ in  \eqref{RT-2point-theta-zero_v1} we obtain 
\eqref{R-2point-theta-zero-v2}. The two-point function in  \eqref{RT-2point-theta-zero_v1} can be written in terms of the thermodynamic quantities $T$ and $\Omega$. The result reads
\begin{eqnarray}   \label{RT-2point-theta-zero_v2}
\frac{1}{\Delta}\log \langle{\cal O}_{\Delta}(0) {\cal O}_{\Delta}(\tau)\rangle &=& - \, 2 \, \log \tau  +\frac{1}{6} \, \Omega^2 \, \tau ^2 - \frac{1}{6} \left(\pi ^2 \, T^2 + \Omega ^2\right) \Omega\, \tau^3
\\[5pt] 
&&+\frac{1}{360} \left(9 \, \pi ^4 \, T^4+18 \, \pi ^2 \, T^2 \, \Omega ^2+13 \, \Omega ^4\right) \tau ^4  - \,\frac{1}{60} 
\left(\pi ^2 \, T^2+\Omega ^2\right) \Omega^3 \, \tau^5 \, . 
\nonumber 
\end{eqnarray}
Considering the large $\Delta$ limit, as specified in \eqref{largeD_limit}, the two-point function in  \eqref{RT-2point-theta-zero_v2} becomes
\begin{equation}   \label{RT-2point-theta-zero_v3}
\langle{\cal O}_{\Delta}(0) {\cal O}_{\Delta}(\tau)\rangle = \frac{1}{\tau^{2 \, \Delta}} \, 
\exp \left[\frac{1}{6} \, \Delta \, \Omega^2 \, \tau ^2  -\frac{\pi^2}{6} \,\Delta \, \Omega \, T^2 \, \tau^3   + \frac{\pi^4}{40} \,\Delta \, T^4 \, \tau^4 \right] \, . 
\end{equation}
Notice the additional term proportional to $\tau^3$ in \eqref{RT-2point-theta-zero_v3} compared to the two-point function of \eqref{R-2point-theta-zero-v3}. We will comment on this in the next subsection where the most general case, in which the two operators have both temporal and spatial boundary distance, will be considered.

\subsubsection{Temporal and spatial boundary distance}
\label{subsubsection:Finite-T-R-temporal-spatial}

Lets us now consider the most general two-point function $\langle{\cal O}_{\Delta}(0) {\cal O}_{\Delta}(\tau, \vec{x})\rangle $, 
for operators that have both temporal and spatial boundary distance. In this case we are looking for geodesics that hit the 
boundary at the points $(0,0)$ and  $(\tau,\vec{x})$ while the meeting point is located by symmetry  at the bulk point $(\tau/2, \vec{x}/2, z_{\max})$. 
In this subsection we will present in detail the computation of the two-point function when the orientation of the string is  
$\theta=\pi/2$ and $\psi=0$, and we refer the reader in appendix \ref{appendix:2point-full} for the relevant details of the 
computation when the string orientation is $\theta=0$ and $\psi=0$.

 \subsubsection*{\underline{$\theta=\pi/2$ and $\psi=0$}}

The expressions for $\dot{t}$ and $\dot x$ in \eqref{tdxd} and the action in \eqref{NGfin}, after substituting 
the metric components from \eqref{metric_comp_theta_pi_over_2}, become
\begin{equation} \label{RT-full-tdot-xdot-S-theta-pi-over2}
\dot t \, = \, \frac{\mu \, z}{\left(1 - \eta^4 \, z^4\right)\sqrt{\big. h(z)}} 
\, , \quad  
\dot x \, = \, \frac{\nu \, z}{\sqrt{\big. h(z)}} 
\quad \& \quad 
S \, = \, \Delta \int \frac{dz}{z} \, \frac{1}{ \left(1- r_0^2 \, z^2\right) \, \sqrt{\big.h(z)}}
\end{equation}
where the function $h(z)$ is defined in this case as follows
\begin{equation}
h(z) = \left(1 - \eta^4 \, z^4\right) \left[ 1- \nu^2 \, z^2 
\left(1- r_0^2 \, z^2\right) \right]- \mu^2 \, z^2 \left(1- r_0^2 \, z^2\right) \, . 
\end{equation}
To determine the turning point of the string we have to impose the vanishing of the function $h(z)$. This equation cannot be 
solved analytically and for this reason we obtain $z_{\max}$ perturbatively for small values of $r_0$ and $\eta$. 
Notice here that if one tries to obtain a perturbative expansion of $z_{\max}$ expanding for large values of $\mu$ and $\nu$, 
then the final expression will not have a smooth limit even  if one tries to isolate the temporal or the spatial distance.  Consequently, the expression for $z_{\max}$ reads
\begin{eqnarray}  \label{RT-full-meeting-point-theta-pi-over2}
&& \left(\mu ^2+\nu ^2\right) z_{\rm max}^2 = 
1+  \frac{r_0^2}{\mu ^2+\nu ^2}+\frac{2 \, r_0^4}{\left(\mu ^2+ \nu ^2\right)^2}  - 
\Bigg[1 + \frac{4 \, r_0^2}{\mu ^2+\nu ^2} +\frac{15\, r_0^4}{\left(\mu ^2+\nu ^2\right)^2}\Bigg] 
\frac{\eta ^4 \, \mu^2}{\left(\mu ^2+\nu ^2\right)^3}
\nonumber \\[5pt]
&& \qquad \qquad 
+\Bigg[\frac{2 \, \mu ^2- \nu ^2}{\mu ^2+\nu ^2}+
\frac{\left(15 \, \mu ^2-6 \, \nu ^2 \right) r_0^2}{\left(\mu ^2+\nu ^2\right)^2}+
\frac{\left(84  \, \mu ^2-28 \,  \nu ^2\right) r_0^4}{\left(\mu ^2+\nu^2\right)^3}\Bigg]  
\frac{\eta ^8 \, \mu^2}{\left(\mu ^2+\nu ^2\right)^5} \, . 
\end{eqnarray}
It is difficult to calculate he integrals in  \eqref{RT-full-tdot-xdot-S-theta-pi-over2}  analytically, so as above we perform the calculations
perturbatively for small values of $r_0$ and $\eta$. For the temporal distance we obtain
\begin{eqnarray}  \label{RT-full-tau-theta-pi-over2}
&& \left(\mu ^2+\nu ^2\right) \frac{\tau}{2 \, \mu} = 1+\frac{4 \, r_0^2}{3 \left(\mu ^2+\nu ^2\right)}+
\frac{16 \, r_0^4}{5 \left(\mu ^2+\nu ^2\right)^2} +
\frac{64 \, r_0^6}{7 \left(\mu ^2+\nu ^2\right)^3} -
\Bigg[\frac{4}{5} \, \frac{\mu ^2-\nu ^2}{\mu ^2+\nu ^2}
\nonumber \\[5pt]
&& \quad 
+
\frac{32 \left(5 \, \mu ^2-3 \, \nu ^2\right) r_0^2}{35 \left(\mu ^2+\nu ^2\right)^2} + 
\frac{64 \left(7 \, \mu ^2-3 \, \nu ^2\right) r_0^4}{21 \left(\mu ^2+\nu^2\right)^3}+ 
\frac{1024 \left(3 \, \mu ^2-\nu ^2\right) r_0^6}{33 \left(\mu ^2+\nu ^2\right)^4} \Bigg] 
\frac{\eta ^4}{\left(\mu ^2+\nu^2\right)^2}
\end{eqnarray}
while for the spatial distance the result is
\begin{eqnarray} \label{RT-full-x-theta-pi-over2}
&& \left(\mu ^2+\nu ^2\right) \frac{x}{2 \, \nu} = 1+
\frac{4 \, r_0^2}{3 \left(\mu ^2+\nu ^2\right)}+
\frac{16 \, r_0^4}{5 \left(\mu ^2+\nu ^2\right)^2} +
\frac{64 \, r_0^6}{7 \left(\mu ^2+\nu ^2\right)^3} -
\Bigg[\frac{4}{15} \, \frac{5\, \mu ^2-\nu ^2}{\mu ^2+\nu ^2}
\nonumber \\[5pt]
&& \quad 
+
\frac{32 \left(7 \, \mu ^2- \nu ^2\right) r_0^2}{35 \left(\mu ^2+\nu ^2\right)^2} + 
\frac{64 \left(9\, \mu ^2- \nu ^2\right) r_0^4}{21 \left(\mu ^2+\nu^2\right)^3}+ 
\frac{1024 \left(11\,  \mu ^2-\nu ^2\right) r_0^6}{99 \left(\mu ^2+\nu ^2\right)^4} \Bigg] 
\frac{\eta ^4}{\left(\mu ^2+\nu^2\right)^2}\, .
\end{eqnarray}
Finally the on-shell action becomes
\begin{eqnarray} \label{RT-full-S-theta-pi-over2}
&& -\frac{2 \, S_{os}}{\Delta} = \log\left[\frac{\mu^2+\nu^2}{4}\right] - 
\frac{4 \, r_0^2}{\mu ^2+\nu ^2}-
\frac{8 \, r_0^4}{\left(\mu ^2+\nu^2\right)^2} - 
\frac{64 \, r_0^6}{3 \left(\mu ^2+\nu ^2\right)^3}+
\Bigg[\frac{2}{3} \, \frac{3\, \mu ^2-\nu ^2}{\mu ^2+\nu ^2}
\nonumber \\[5pt]
&& \quad 
+
\frac{32 \left(5 \, \mu ^2- \nu ^2\right) r_0^2}{15 \left(\mu ^2+\nu ^2\right)^2} + 
\frac{48 \left(7\, \mu ^2- \nu ^2\right) r_0^4}{7 \left(\mu ^2+\nu^2\right)^3}+ 
\frac{1024 \left(9\,  \mu ^2-\nu ^2\right) r_0^6}{45 \left(\mu ^2+\nu ^2\right)^4} \Bigg] 
\frac{\eta ^4}{\left(\mu ^2+\nu^2\right)^2} \, . 
\end{eqnarray}
We also introduce the Euclidean distance of the two operators, as well as the ratio  of $\tau$ to $|x|$
\begin{equation} \label{def-gamma-xi}
|x| = \sqrt{\big.\tau^2 + x^2} \quad \& \quad \xi = \frac{\tau}{|x|} \, .
\end{equation}
Combining the expansions \eqref{RT-full-tau-theta-pi-over2}, \eqref{RT-full-x-theta-pi-over2} and \eqref{RT-full-S-theta-pi-over2} together with the definition of $|x|$ and $\xi$ from \eqref{def-gamma-xi}, we obtain the following expansion 
of the two-point function for small values of $\eta$ and $r_0$
\begin{eqnarray}  \label{RT-full-2point-theta-pi-over2_v1}
&& \frac{1}{\Delta}\log \langle{\cal O}_{\Delta}(0) {\cal O}_{\Delta}(\tau, \vec{x})\rangle = 
- \, 2 \, \log |x|  -\frac{1}{3} \, r_0^2 \, |x|^2  +\frac{1}{90} \, r_0^4 \, |x|^4- \frac{2}{2835} \, r_0^6 |x|^6 
\\[5pt]
&& \qquad + \,
\frac{1}{120} \left(4\, \xi^2 -1\right)  \eta^4 \, |x|^4 
- \frac{1}{315} \left(5\, \xi^2 -2\right) r_0^2 \, \eta^4\, |x|^6 
+ \frac{1}{8400} \left(56\, \xi^2 -29\right) r_0^4 \, \eta^4\, |x|^8 \, .
\nonumber
\end{eqnarray}
Notice that setting $\xi=1$ in  \eqref{RT-full-2point-theta-pi-over2_v1} we obtain the expression for the two-point function in 
\eqref{RT-2point-theta-pi-over2_v1}. It is useful to express the two-point function in  \eqref{RT-full-2point-theta-pi-over2_v1}  in terms of the thermodynamic quantities 
$T$ and $\Omega$. The result reads
\begin{eqnarray}   \label{RT-full-2point-theta-pi-over2_v2}
&& \frac{1}{\Delta}\log \langle{\cal O}_{\Delta}(0) {\cal O}_{\Delta}(\tau,\vec{x})\rangle = - \, 2 \, \log |x|  -\frac{1}{3} \, \Omega^2 \, |x| ^2 
\nonumber \\[5pt] 
&&\qquad \qquad \qquad \qquad + \, \frac{1}{360} \Bigg[4 \, \Omega^4+ 3 \left(4\, \xi^2-1\right) 
\left[(\Omega^2 + \pi^2 \, T^2  \right)^2 \Bigg] |x|^4 
\\[5pt] 
&&\qquad \qquad \qquad \qquad  - \,  \,\frac{1}{2835} \, \Omega^2 \, \Bigg[ 2 \, \Omega^4+9\left(5\, \xi^2-2\right)  \left(\Omega^2 + \pi^2 \, T^2  \right)^2\Bigg] |x|^6 
\nonumber \\[5pt] 
&&\qquad \qquad \qquad \qquad  
+ \, \frac{1}{8400} \left(56\, \xi^2 -29\right) \Omega^4 \,  \left(\Omega^2 + \pi^2 \, T^2  \right)^2 \, |x|^8 \, . 
\nonumber 
\end{eqnarray}
Notice that setting $\xi=1$ in  \eqref{RT-full-2point-theta-pi-over2_v2} we obtain the expression for the two-point function in 
\eqref{RT-2point-theta-pi-over2_v2}. Considering the large $\Delta$ limit of  \eqref{largeD_limit} (in which one should replace $\tau$ by $|x|$), the two-point function in  \eqref{RT-full-2point-theta-pi-over2_v2} becomes
\begin{equation}    \label{RT-full-2point-theta-pi-over2_v3}
\langle{\cal O}_{\Delta}(0) {\cal O}_{\Delta}(\tau, \vec{x})\rangle = \frac{1}{|x|^{2 \, \Delta}} \, 
\exp \left[- \frac{1}{3} \, \Delta \, \Omega^2 \, |x|^2 + \frac{\pi^4}{120}\, C_2^{(1)}(\xi) \,\Delta \, T^4 \, |x|^4 \right] \, 
\end{equation}
where $C_2^{(1)}(\xi)$ is the following Gegenbauer polynomial
\begin{equation}
C_2^{(1)}(\xi) = 4 \, \xi^2 - 1 = \frac{3 \, \tau^2 -\vec{x}^2}{\tau^2 +\vec{x}^2} \, . 
\end{equation}
Furthermore in the large $\Delta$ limit that we consider the two-point function can be resummed into an exponential of the blocks of the energy-momentum tensor and the scalar operator $\phi^2$.

 \subsubsection*{\underline{$\theta=0$ and $\psi=0$}}

When the orientation of the string is  $\theta=0$ and $\psi=0$, the expression for the two-point function as an expansion 
for small values of $\eta$ and $r_0$ becomes
\begin{eqnarray}  \label{RT-full-2point-theta-zero_v1}
&& \frac{1}{\Delta}\log \langle{\cal O}_{\Delta}(0) {\cal O}_{\Delta}(\tau, \vec{x})\rangle = 
- \, 2 \, \log |x|  
+ \frac{1}{6} \, r_0^2 \, |x|^2 
- \frac{1}{6} \, \xi \, r_0 \, \eta^2  \, |x|^3 
+ \frac{1}{90} \, r_0^4 \, |x|^4 
- \frac{1}{60} \, \xi \, r_0^3 \, \eta^2 \, |x|^5
\nonumber \\[5pt]
&& \qquad \qquad 
+ \frac{1}{120} \, \left(4 \, \xi^2 -1 \right) \eta^4 \, |x|^4 
+ \frac{1}{1680}  \, \left(34 \, \xi^2 -1 \right) r_0^2\, \eta^4 \,  |x|^6 
- \frac{1}{210} \, \xi \, r_0^5 \, \eta^2 \, |x|^7 \, . 
\end{eqnarray}
The two-point function in  \eqref{RT-full-2point-theta-zero_v1} can be written in terms of the thermodynamic quantities 
$T$ and $\Omega$. The result reads
\begin{eqnarray}   \label{RT-full-2point-theta-zero_v2}
&& \frac{1}{\Delta}\log \langle{\cal O}_{\Delta}(0) {\cal O}_{\Delta}(\tau, \vec{x})\rangle = - \, 2 \, \log \tau  +\frac{1}{6} \, \Omega^2 \, |x| ^2 - \frac{1}{6} \left(\pi ^2 \, T^2 + \Omega ^2\right) \Omega\, \xi \, |x|^3
\nonumber \\[5pt] 
&&\quad \quad \quad  +\frac{1}{360} \Bigg[ 4 \, \Omega^4 + 3  \left(4 \, \xi^2-1\right) \left(\Omega^2 + \pi^2 \, T^2 \right)^2 \Bigg] |x| ^4  
 - \frac{1}{60} \, \xi \, \Omega^3 \, \left(\Omega^2 + \pi^2 \, T^2 \right) \, |x|^5
\nonumber \\[5pt] 
&& \quad \quad \quad
+ \frac{1}{1680}  \, \left(34 \, \xi^2 -1 \right) \Omega^2\, \left(\Omega^2 + \pi^2 \, T^2 \right)^2 \,  |x|^6 
- \frac{1}{210} \, \xi \, \Omega^5 \, \left(\Omega^2 + \pi^2 \, T^2 \right) \, |x|^7\, . 
\end{eqnarray}
Notice that setting $\xi=1$ in  \eqref{RT-full-2point-theta-zero_v2} we obtain the expression for the two-point function in 
\eqref{RT-2point-theta-zero_v2}. 
Considering the large $\Delta$ limit of  \eqref{largeD_limit}, the two-point function in  \eqref{RT-full-2point-theta-zero_v2} becomes
\begin{equation}    \label{RT-full-2point-theta-zero_v3}
\langle{\cal O}_{\Delta}(0) {\cal O}_{\Delta}(\tau, \vec{x})\rangle = \frac{1}{|x|^{2 \, \Delta}} \, 
\exp \left[\frac{1}{6} \, \Delta \, \Omega^2 \, |x|^2 -\frac{\pi^2}{12}\, C_1^{(1)} (\xi)\, \Delta \, \Omega\, T^2 \, |x|^3+ \frac{\pi^4}{120}\, C_2^{(1)}(\xi) \,\Delta \, T^4 \, |x|^4 \right] 
\end{equation}
where $C_1^{(1)}(\xi)$ is the following Gegenbauer polynomial
\begin{equation}
C_1^{(1)}(\xi) = 2 \, \xi\, . 
\end{equation}
When expanding the exponential of \eqref{RT-full-2point-theta-zero_v3} the three leading terms in the expansion are associated to the non-zero expectation value that the scalar operator $\phi^2$, the R-current ${\cal J}_{\phi_3}^{\mu}$ and the stress energy tensor $T^{\mu\nu}$, respectively acquire. Furthermore in the large $\Delta$ limit that we consider the two-point function can be resummed into an exponential of the blocks of the energy-momentum tensor, the R-current and the scalar operator $\phi^2$.

In what follows, we will show that the two-point function \eqref{RT-full-2point-theta-zero_v3}  can actually be written in the form dictated by the OPE, that is in the form \eqref{2-point-R}. To this end we rewrite the last term in the exponent of \eqref{RT-full-2point-theta-zero_v3} as a linear combination of Gegenbauer polynomials, as was also done in \cite{Rodriguez-Gomez:2021mkk}.
\begin{equation}\label{RT-5}
G_2 \, = \frac{1}{|x|^{2 \, \Delta}} \, 
\exp \left[\frac{1}{6} \, \Delta \, \Omega^2 \, |x|^2 +i \,  a \, \xi \right] \sum_{J=0}\sum_{n=0}{1 \ov n!}
\left({\pi^4 \Delta \ov120} \right)^n {{\cal C}_{J,n} \ov \beta^{4 n}} |x|^{4 \, n} C_J^{(1)}(\xi) 
\end{equation}
where for notational convenience we have defined 
\begin{equation}\label{a}
a=i \, {\pi^2 \ov 6} \, \Delta \, \Omega \, T^2 \, |x|^3
\end{equation}
and the coefficients ${\cal C}_{J,n}$ are given by \cite{Rodriguez-Gomez:2021mkk}
\begin{equation}
\label{Cs}
\mathcal{C}_{J,n}=\frac{2}{\pi}\,\int_{-1}^1d\eta\,\sqrt{1-\eta^2}\,C_J^{(1)}(\eta)\,\big(C_2^{(1)}(\eta)\big)^n=\begin{cases}C_{n-\frac{J}{2}}^{(-n)}(-\frac{1}{2})-C_{n-\frac{J}{2}-1}^{(-n)}(-\frac{1}{2}) \quad 
&{\rm if}\ n,\,J\ne 0\\ 1\quad &{\rm if}\ n,\,J=0\end{cases}
\end{equation}
In order to proceed one should express $e^{i a \xi}$ as a linear combination of the Gegenbauer polynomials
\begin{eqnarray}\label{axi}
&& \qquad \qquad \qquad \qquad 
e^{i a \xi}=\sum_{\tilde J=0} {\cal D}_{\tilde J}(a) \, C_{\tilde J}^{(1)}(\xi) \quad \Rightarrow  \quad 
\\
&& \int_{-1}^1d\xi \sqrt{1-\xi^2}\, e^{i \, a \, \xi}\,  C_{\hat J}^{(1)}(\xi)=\sum_{\tilde J=0}{\cal D}_{\tilde J}(a) \int_{-1}^1d\xi \sqrt{1-\xi^2}\, C_{\tilde J}^{(1)}(\xi)\, C_{\hat J}^{(1)}(\xi) \, . 
\nonumber
\end{eqnarray}
For the left hand side of the equation above one uses identity in the first line of \eqref{idss}, while for the right hand side the orthogonality 
of the Gegenbauer polynomials which is at the second line of \eqref{idss}
\begin{eqnarray}\label{idss}
&& \int_{-1}^1d\xi \sqrt{1-\xi^2}\, e^{i \, a \, \xi}\,  C_{\hat J}^{(1)}(\xi)={\pi \, i^{\hat J} \, \Gamma(2+\hat J)\ov \hat J!} \, 
a^{-1} J_{1+\hat J}(a)
\nonumber \\
&& \int_{-1}^1d\xi \sqrt{1-\xi^2}\, C_{\tilde J}^{(1)}(\xi)\, C_{\hat J}^{(1)}(\xi)=\delta_{J \hat J} \, 
{\pi \, \Gamma(2+\hat J)\ov 2 \, \hat J \, !(\hat J+1)} \, . 
\end{eqnarray}
Consequently, \eqref{axi} leads to
\begin{equation}
{\cal D}_{\hat J}(a) \, = \, 2 \, \left(\hat J+1\right) \, i^{\hat J} \, a^{-1} \, J_{{\hat J} +1}(a) 
\end{equation}
where $J_{{\hat J}+1}(a)$ is the Bessel function of the first kind and admits the following expansion 
\begin{equation}\label{Bessel}
J_{\nu}(a)=\sum_{l=0}^\infty{(-1)^l \ov l! \left(l+\nu+1\right)}\left({a \ov 2} \right)^{2 \, l+\nu} \, . 
\end{equation}
Finally, taking into account the completeness relation for the Gegenbauer polynomials
\begin{equation}
C_{J}^{(1)}(\xi)\, C_{\hat J}^{(1)}(\xi) \, = \, \sum_{K=0}^{{\rm min} \left(J, \hat J\right) } C_{|J-\hat J|+2\,K}^{(1)}(\xi)
\end{equation}
we get for the 2-point correlator 
\begin{eqnarray}\label{RT-6}
&&G_2= \frac{1}{|x|^{2 \, \Delta}} \, 
\exp \left[\frac{1}{6} \, \Delta \, \Omega^2 \, |x|^2  \right] \times \\
&&\sum_{J=0}\sum_{n=0}\sum_{\hat J=0} \sum_{K=0}^{{\rm min} \left(J, \hat J\right)}{2\, i^{\hat J} \ov n!}
\left({\pi^4 \Delta \ov120 } \right)^n  \, {{\cal C}_{J,n} \ov \beta^{4 n}}  \, |x|^{4n} \left(\hat J+1\right) \,  a^{-1}  \, 
J_{1+\hat J}(a) \, C_{|J-\hat J|+2\,K}^{(1)}(\xi)
\nonumber
\end{eqnarray}
where $a$ is given by \eqref{a}. Equation \eqref{RT-6} should be written in terms of powers of $|x|$ the temperature $T$ and the chemical potential $\Omega$ by employing \eqref{Bessel}. The same also holds for the first line of \eqref{RT-6}.
It is evident now that the holographic result \eqref{RT-6} takes the form dictated by the OPE \eqref{2-point-R}.
Each term of the form $|x|^{\Delta'}C_{J'}^{(1)}(\xi)$ in \eqref{RT-6} represents the contribution in the OPE of an operator of dimension $\Delta'$ and spin $J'$.


\section{Conclusions}

The aim of this paper is to study conformal field theories in the presence of finite density and/or finite temperature. In particular, we considered ${\cal N}=4$ SYM at finite density and/or finite temperature. The dual string theory background \eqref{metric-general-v1} is generated by non-extremal rotating D3-branes with two equal non-zero angular momenta. The presence of the metric elements $g_{t\phi_{2,3}}$, that connect the internal space to the asymptotically $AdS_5$ space, perplexes the holographic calculation of correlators in this kind of geometries. We have developed a general method for calculating correlation functions of operators of large dimension,
that are dual to classical string states at strong coupling, by consistently integrating out the coordinates of the internal space. In this way, 
the resulting effective action \eqref{NG} describes the motion of a point-like string in the 5-dimensional asymptotically $AdS_5$ space.

Subsequently, we studied the OPE of two scalar operators in the dual field theory. In the case of zero temperature but non-zero R-charge density the geometry introduces a scale $r_0$.  This, in turn, implies that some 
of the field theory operators develop a VEV. However, we argue that this can happen only for scalar operators, that is all operators with spin have zero expectation value. As a result, the OPE dictates that the two-point correlator can be written as a sum over only scalar fields \eqref{2-point-R}. Each term in the sum is proportional to $r_0^{\Delta}$, where $\Delta$ is the dimension of the scalar operator whose expectation value is non-zero. The constant of proportionality depends crucially on the CFT data, and more precisely on the fusion coefficients of the underlying CFT. Then by using the aforementioned method we have calculated holographically the one-point function of scalar operators of large dimension. The calculation boils down to finding the length of a geodesic of a particle travelling  from the point of the boundary, where the dual field theory operator is situated, all the way down to the apparent singularity of the geometry. The result is in agreement with the structure of the OPE. Next, we holographically evaluated the two-point function for operators whose string duals sit at two different points of the internal space as functions of the scale $r_0$. In a particular large $\Delta$ limit  the result for the two-point correlators exponentiate (see \eqref{R-2point-theta-pi-over2-v3} and \eqref{R-2point-theta-zero-v3}). Finally, let us mention that in the limit of zero R-charge $r_0\rightarrow 0$ one smootly obtains the results for the one and two-point functions in $AdS_5$.

In the next section \ref{section:Finite-T}, we considered the background with non-zero temperature but  with the chemical potential being zero. We focused on the holographic calculation of two-point correlators whose temporal or spatial distance is very large. In the case of large temporal distance the periodicity of the time coordinate sets an upper bound (which is half of the period, i.e. ${1\ov 2 T}$) to the distance of the two operators. The result of the two-point function in this limit is given by \eqref{T-tau-2point-perturbative-large-eta}. Subsequently, we calculated the two-point function in the limit of large spatial distance. In this case we found that the correlators fall exponentially with the distance  \eqref{G2}, as one would have expected.

In the presence of finite temperature and finite density operators with and without spin can have non-zero expectation value \eqref{1-point-2}. 
The general form of the OPE can be used to write the two-point correlators as an infinite sum, where each of the terms in the sum is related to expectation value of an operator of definite dimension and spin \eqref{2-point-RT}. As in the case of finite temperature,
 one may write the two-point function as a linear combination of the Gegenbauer polynomials $C_J^{(1)}(\xi)$ but with the coefficients depending now on both the temperature and the 
chemical potential, as well as on the CFT data. 
We present a systematic expansion of the holographic two-point correlators in powers of the temperature T and of the chemical potential $\Omega$. For a string sitting at $\theta=\pi/2,\, \psi=0$ the leading terms in the expansion \eqref{RT-full-2point-theta-pi-over2_v3} originate from the expectation values of the scalar operator $\phi^2$ and the stress-energy tensor $T^{\mu\nu}$,
while for the two-point correlator dual to the string sitting at $\theta=0,\, \psi=0$ 
the leading terms in the expansion \eqref{RT-full-2point-theta-zero_v3} originate from the expectation values of the scalar operator $\phi^2$, the R-current ${\cal J}^\mu_{\phi_3}$ and the stress-energy tensor $T^{\mu\nu}$.
By employing the Ward identity for the R-current and by comparing the appropriate term of the holographic result for the two-point correlator to the corresponding term in the OPE we derived the value 
of the R-charge density of the background and found compelling agreement with the analysis of thermodynamics of the black hole. Finally, in a certain large $\Delta$ limit (see \eqref{largeD_limit}) the result for the two-point correlators exponentiate (see \eqref{RT-full-2point-theta-pi-over2_v3} and \eqref{RT-full-2point-theta-zero_v3}). 
One may expand the exponential to show that the result takes precisely the form dictated by the OPE, that is it is written in terms of the Gegenbauer polynomials \eqref{RT-6} with each term representing the contribution of an operator with definite dimension and spin.
Finally, let us mention that in the limit of zero R-charge or zero temperature one smoothly obtains the results for the one and two-point functions of the previous sections. 

An interesting future direction could be the generalisation of the method presented in section \ref{section:backgrounds} 
to the case in which the string solution involves more than one $U(1)$ isometries, as well as to the case of extended strings. 
Also interesting could be to calculate holographically higher point correlation functions. 
Finally, it would be important to specify the form of the corrections to the type-IIB supergravity action which will reproduce the one-point functions in the presence of temperature and chemical potential. 
We conjecture that these will be products of the Weyl tensor with the five-form field. However, such a computation seems extremely complicated since the geometry near the singularity breaks down.


\section*{Acknowledgments}

The research work of this project was supported by the Hellenic Foundation
for Research and Innovation (H.F.R.I.) under the “First Call for
H.F.R.I. Research
Projects to support Faculty members and Researchers and the procurement of
high-cost research equipment grant” (MIS 1857, Project Number: 16519).

\appendix

\section{Two-point function with temporal/spatial dependence and finite T/R}
\label{appendix:2point-full}

In this appendix we provide all the intermediate steps in order to reproduce equation  \eqref{RT-full-2point-theta-zero_v1}.

The expressions for $\dot{t}$ and $\dot x$ in \eqref{tdxd} and the action in \eqref{NGfin}, after substituting 
the metric components from \eqref{metric_comp_theta_zero}, become
\begin{eqnarray}  \label{RT-full-tdot-xdot-S-theta-zero}
&& \dot t \, = \, \frac{\mu \, z - r_0 \, \eta^2 \, z^3}{\left(1 - \eta^4 \, z^4\right)\sqrt{\big. \left(1 - r_0^2 \, z^2\right)h(z) }} 
\, , \quad  
\dot x \, = \, \frac{\nu \, z}{\sqrt{\big. \left(1 - r_0^2 \, z^2\right)h(z)}} 
\nonumber \\[5pt]
&& \qquad \qquad 
S \, = \, \Delta \int \frac{dz}{z} \,\frac{1 - r_0^2 \, z^2 - \eta^4 \, z^4 +r_0 \, \eta^2 \, \mu \, z^4}{\left(1 - \eta^4 \, z^4\right)\sqrt{\big. \left(1 - r_0^2 \, z^2\right)h(z) }}
\end{eqnarray}
and the function $h(z)$ is defined in this case as follows
\begin{equation}
h(z) =1 -\eta^4 \, z^4\left(1 - \nu^2 \, z^2 \right) - \left(\mu^2+ \nu^2 + r_0^2 \right) z^2 +2 \, r_0 \, \eta^2 \, \mu \, z^4 \, . 
\end{equation}
From the vanishing of the function $h(z)$ we determine the $z$ coordinate for the turning point of the string, and it is 
\begin{eqnarray}
&& \left(\mu ^2+\nu ^2\right) z_{\rm max}^2 = 
1-  \frac{r_0^2}{\mu ^2+\nu ^2}+\frac{ r_0^4}{\left(\mu ^2+ \nu ^2\right)^2}  + 
2 \Bigg[1 - \frac{3 \, r_0^2}{\mu ^2+\nu ^2} \Bigg] 
\frac{\mu \, r_0\, \eta^2}{\left(\mu ^2+\nu ^2\right)^2}
\nonumber \\[5pt]
&& \qquad \qquad 
- \Bigg[\frac{\mu ^2}{\mu ^2+\nu ^2} - \frac{\left(11 \, \mu ^2- \nu ^2\right) \, r_0^2}{\left(\mu ^2+\nu ^2\right)^2} +
\frac{ \left(46 \, \mu ^2- 4\, \nu ^2\right)  r_0^4}{\left(\mu ^2+\nu ^2\right)^3}\Bigg] 
\frac{\eta ^4}{\left(\mu ^2+\nu ^2\right)^2} \, .
\end{eqnarray}
The calculation of the integrals in  \eqref{RT-full-tdot-xdot-S-theta-zero} will determine the temporal boundary distance
\begin{eqnarray}  \label{RT-full-tau-theta-zero}
&& \left(\mu ^2+\nu ^2\right) \frac{\tau}{2 \, \mu} = 1-\frac{2 \, r_0^2}{3 \left(\mu ^2+\nu ^2\right)}+
\frac{8 \, r_0^4}{15 \left(\mu ^2+\nu ^2\right)^2} +
\Bigg[ \frac{3\, \mu ^2-\nu ^2}{\mu ^2+\nu ^2}-
\frac{8 \left(5 \, \mu ^2-\nu ^2\right) r_0^2}{5 \left(\mu ^2+\nu ^2\right)^2} \Bigg] \frac{2\, r_0 \, \eta^2}{3\, \mu \left(\mu ^2+\nu ^2\right)}
\nonumber \\[5pt]
&& \quad -
\Bigg[\frac{4}{5} \, \frac{\mu ^2-\nu ^2}{\mu ^2+\nu ^2}-
\frac{16 \left(5 \, \mu ^2-3 \, \nu ^2\right) r_0^2}{7 \left(\mu ^2+\nu ^2\right)^2} + 
\frac{704 \left(7 \, \mu ^2-3 \, \nu ^2\right) r_0^4}{105 \left(\mu ^2+\nu^2\right)^3}\Bigg] 
\frac{\eta ^4}{\left(\mu ^2+\nu^2\right)^2}\, ,
\end{eqnarray}
the spatial boundary distance
\begin{eqnarray}  \label{RT-full-xi-theta-zero}
&& \left(\mu ^2+\nu ^2\right) \frac{x}{2 \, \nu} = 1-\frac{2 \, r_0^2}{3 \left(\mu ^2+\nu ^2\right)}+
\frac{8 \, r_0^4}{15 \left(\mu ^2+\nu ^2\right)^2} +
\Bigg[ 1- \frac{12 \, r_0^2}{5 \left(\mu ^2+\nu ^2\right)} \Bigg] \frac{8\, r_0 \, \mu  \, \eta^2}{3 \left(\mu ^2+\nu ^2\right)^2}
\nonumber \\[5pt]
&& \quad -
\Bigg[\frac{4}{15} \, \frac{5\, \mu ^2-\nu ^2}{\mu ^2+\nu ^2}-
\frac{16 \left(7 \, \mu ^2- \nu ^2\right) r_0^2}{7 \left(\mu ^2+\nu ^2\right)^2} + 
\frac{704 \left( 9\, \mu ^2- \nu ^2\right) r_0^4}{105 \left(\mu ^2+\nu^2\right)^3}\Bigg] 
\frac{\eta ^4}{\left(\mu ^2+\nu^2\right)^2}\, ,
\end{eqnarray}
as well as, the on-shell action
\begin{eqnarray}  \label{RT-full-S-theta-zero}
&& -\frac{2 \, S_{os}}{\Delta} = \log\left[\frac{\mu^2+\nu^2}{4}\right] +
\frac{2 \, r_0^2}{\mu ^2+\nu ^2}-
\frac{4 \, r_0^4}{3\left(\mu ^2+\nu^2\right)^2} -
\Bigg[ 1- \frac{12 \, r_0^2}{5 \left(\mu ^2+\nu ^2\right)} \Bigg] \frac{16\, r_0 \, \mu  \, \eta^2}{3 \left(\mu ^2+\nu ^2\right)^2}
\nonumber \\[5pt]
&& \quad 
+ \Bigg[\frac{2}{3} \, \frac{3\, \mu ^2-\nu ^2}{\mu ^2+\nu ^2}-
\frac{16 \left(5 \, \mu ^2- \nu ^2\right) r_0^2}{3 \left(\mu ^2+\nu ^2\right)^2} + 
\frac{528 \left(7\, \mu ^2- \nu ^2\right) r_0^4}{35 \left(\mu ^2+\nu^2\right)^3} \Bigg] 
\frac{\eta ^4}{\left(\mu ^2+\nu^2\right)^2} \, . 
\end{eqnarray}
Combining  \eqref{RT-full-tau-theta-zero},  \eqref{RT-full-xi-theta-zero} and  \eqref{RT-full-S-theta-zero}, it is easy to obtain the 
equation \eqref{RT-full-2point-theta-zero_v1} of the main text.


\end{document}